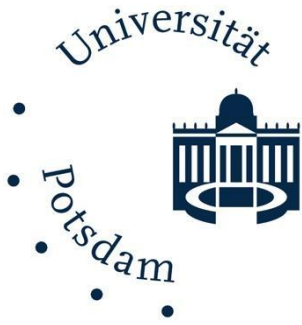

**Universität Potsdam**

Institut für Informatik und Computational Science

Lehrstuhl für Software Engineering

**Bachelorarbeit**

Zur Erlangung des akademischen Grades Bachelor of Science (B. Sc.)

# Evaluierung der Code-Generierungsfähigkeiten von ChatGPT 4: Eine vergleichende Analyse in 19 Programmiersprachen

Von Laurenz Gilbert
Matrikelnummer: 808291
Abgabedatum: 18.06.2024
Studiengang: Wirtschaftsinformatik

Erstprüferin: **Prof. Dr. Anna-Lena Lamprecht**
Zweitprüfer: **PD Dr. Henning Bordihn**

# Abstract

Diese Bachelorarbeit untersucht die Fähigkeiten von ChatGPT 4 zur Code-Generierung in 19 Programmiersprachen. Betrachtet wurden die Lösungsraten zwischen drei Schwierigkeitsgraden, die aufgetretenen Fehlerarten und die Qualität des Codes hinsichtlich der Laufzeit- und Speichereffizienz in einem quantitativen Experiment. Dabei wurden 188 Programmierprobleme der Plattform LeetCode entnommen, wobei ChatGPT 4 jeweils drei Versuche hatte, mittels Feedback eine korrekte Lösung zu generieren. ChatGPT 4 löste 39,67 % aller Aufgaben erfolgreich, wobei die Erfolgsrate mit zunehmendem Schwierigkeitsgrad deutlich abnahm und bei komplexen Problemen in allen Sprachen signifikante Schwierigkeiten auftraten. Das Modell zeigte eine höhere Kompetenz in weit verbreiteten Sprachen, was wahrscheinlich auf eine größere Menge und höhere Qualität der Trainingsdaten zurückzuführen ist. Bezüglich der Lösungsraten zeigte das Modell zudem eine Präferenz für Sprachen mit niedrigem Abstraktionsniveau und statischer Typisierung. Bei Sprachen hoher Popularität trat der Fehler *Wrong Answer* am häufigsten auf, während bei weniger populären Sprachen Compiler- und Laufzeitfehler überwogen, was auf häufige Missverständnisse und Verwechslungen bezüglich der spezifischen strukturellen Eigenschaften dieser Sprachen zurückzuführen ist. ChatGPT 4 demonstrierte in allen Programmiersprachen eine überdurchschnittliche Laufzeiteffizienz und tendierte diesbezüglich erneut zu statisch typisierten und niedrig abstrahierten Sprachen. Die Werte zur Speichereffizienz variierten erheblich, wobei in 14 Sprachen überdurchschnittliche und in fünf Sprachen unterdurchschnittliche Werte erzielt wurden. Es zeigte sich diesbezüglich eine leichte Tendenz zugunsten von niedrig abstrahierten sowie eine Präferenz zu dynamisch typisierten Sprachen. Zukünftige Forschung sollte eine höhere Anzahl an Aufgaben, Iterationen und unpopulären Sprachen einbeziehen. Darüber hinaus könnten die Fähigkeiten von ChatGPT 4 in der Code-Interpretation und -Zusammenfassung, im Debugging und in der Entwicklung komplexer, praxisbezogener Codes analysiert werden.

# Inhaltsverzeichnis

# Abbildungsverzeichnis

# Tabellenverzeichnis

# 1 Einleitung

Mit der Veröffentlichung von ChatGPT 4 im März 2023 hat OpenAI neue Maßstäbe in der Entwicklung künstlicher Intelligenz gesetzt (OpenAI et al., 2024). Im Vergleich zum Vorgängermodell ChatGPT 3.5 weist die neueste Version der GPT-Reihe signifikant erweiterte Kenntnisse in verschiedenen Bereichen auf, darunter Rechts- und Naturwissenschaften, Sprachen, Mathematik und Medizin. Diese Bachelorarbeit widmet sich einem spezifischen Bereich, in dem ChatGPT 4 erweiterte Kompetenzen demonstriert: der Generierung von Programmcode (OpenAI et al., 2024).

In aktuellen Benchmarks zur Leistungsfähigkeit von Large Language Models (LLMs) bezüglich der Lösung von Programmieraufgaben demonstrieren die GPT-4-Modelle nicht nur eine signifikant bessere Performance im Vergleich zu GPT-3.5, sondern erreichen auch im Vergleich zu anderen populären LLMs Spitzenpositionen (Liu et al., 2023; *EvalPlus Leaderboard*, o. D.). Es fällt jedoch auf, dass sich die gängigen Benchmarks in der Regel auf eine oder wenige Programmiersprachen beschränken und die Ergebnisse als Maß für die allgemeinen Code-Generierungsfähigkeiten der Large Language Models interpretiert werden (Chen et al., 2021; Liu et al., 2023; Yu et al., 2024; Lai et al., 2023). Ein umfassender Vergleich der Leistung von ChatGPT 4 zwischen vielen verschiedenen Programmiersprachen fehlt derzeit, wodurch potenzielle Unterschiede zwischen den Sprachen bislang unberücksichtigt bleiben. Die vorliegende Bachelorarbeit zielt darauf ab, zur Schließung dieser Forschungslücke beizutragen, indem die Leistungsdifferenzen von ChatGPT 4 zwischen 19 ausgewählten Programmiersprachen bei der Lösung allgemeiner Programmierprobleme systematisch untersucht werden.

Die Analyse umfasst die Betrachtung der Lösungsraten, differenziert nach Schwierigkeitsgrad der Aufgaben, eine detaillierte Analyse der aufgetretenen Fehler sowie eine Bewertung der Codequalität anhand der Laufzeit- und Speicherwerte im Vergleich zu den Werten der Lösungen anderer Nutzer. Zur Strukturierung dieser Untersuchung werden drei Forschungsfragen formuliert:

Forschungsfrage 1 (F1): *Wie effektiv löst ChatGPT 4 allgemeine Programmierprobleme unterschiedlicher Schwierigkeitsgrade in 19 verschiedenen Programmiersprachen?*

Forschungsfrage 2 (F2): *Welche Fehler treten in den von ChatGPT 4 in 19 verschiedenen Programmiersprachen generierten Lösungen auf?*

Forschungsfrage 3 (F3): *Wie laufzeit- und speichereffizient sind die von ChatGPT 4 in 19 verschiedenen Programmiersprachen generierten Lösungen?*

Zur Beantwortung der Forschungsfragen wird ChatGPT 4 in 19 ausgewählten Programmiersprachen auf 188 Code-Herausforderungen der Coding-Interview-Plattform LeetCode getestet. Dabei erhält ChatGPT 4 drei Versuche, in denen der Code mithilfe von Feedback verbessert werden soll. Alle Daten werden sorgfältig in einem Git-Repository dokumentiert und anschließend einer deskriptiven statistischen Analyse unterzogen, um Trends und Cluster zu ermitteln, die zur Beantwortung der Forschungsfragen dienen. Um eine solch umfassende Untersuchung zu ermöglichen, wird die gesamte Datenerhebung und -analyse durch eigens für diese Arbeit entwickelte Python-Skripte automatisiert.

Die Bachelorarbeit ist wie folgt strukturiert: In Abschnitt 2, dem theoretischen Hintergrund, werden grundlegende Begriffe definiert und der aktuelle Forschungsstand erörtert. Abschnitt 3 widmet sich der Methodik der Arbeit. Dieser Abschnitt umfasst sowohl die Darstellung der durchgeführten Literaturrecherche zum aktuellen Forschungsstand als auch eine detaillierte Beschreibung des quantitativen Experiments. Dazu wird ein Überblick über die ausgewählten Programmiersprachen und Tools gegeben sowie die Implementierung des Python-Projekts präzise erläutert. Die Ergebnisse der Studie werden in Abschnitt 4 thematisch nach den Forschungsfragen geordnet präsentiert. In Abschnitt 5 erfolgt eine Diskussion dieser Ergebnisse, wobei Implikationen, Limitationen und Ansätze für zukünftige Forschungen berücksichtigt werden. Abschließend wird in Abschnitt 6 die Arbeit mit einem Fazit zusammengefasst.

# 2 Theoretischer Hintergrund

Im folgenden Abschnitt werden einerseits zentrale Begriffe, die für das Verständnis dieser Arbeit essenziell sind, definiert und andererseits der aktuelle Forschungsstand zu den Code-Generierungsfähigkeiten von ChatGPT 4 in verschiedenen Programmiersprachen erörtert.

## 2.1 Begriffsdefinitionen

Die Begriffsdefinitionen umfassen erstens das in dieser Bachelorarbeit verwendete KI-Modell ChatGPT 4, zweitens eine Erläuterung zu den sogenannten Large Language Models (LLMs) und drittens eine Beschreibung der Plattform LeetCode, welche die Quelle der in dieser Untersuchung verwendeten Aufgaben darstellt und auf der die Lösungen von ChatGPT 4 bewertet werden.

### 2.1.1 ChatGPT 4

ChatGPT 4 ist ein interaktiver Chatbot, der auf der im März 2023 von OpenAI vorgestellten GPT-4-Architektur basiert und im Vergleich zum Vorgängermodell, ChatGPT 3.5, weitreichend überlegene Fähigkeiten aufweist, beispielsweise in der Generierung von Programmcode (OpenAI et al., 2024; OpenAI, o. D.-a). Während des sogenannten *Pre-Trainings* wurde GPT-4 auf einem umfangreichen Datenkorpus, der sowohl öffentlich zugängliche als auch von OpenAI lizenzierte Daten umfasst, trainiert und anschließend durch *Reinforcement Learning from Human Feedback* (RLHF) feinabgestimmt (OpenAI et al., 2024). Darüber hinaus wird das Modell kontinuierlich durch Nutzerdaten, beispielsweise durch Interaktionen mit ChatGPT 4, weiter trainiert (OpenAI, o. D.-b). Dank dieses Trainingsprozesses ist das Modell in der Lage, das nächste Wort eines Textes anhand von Gewichtungen, den sogenannten *weights*, präzise vorherzusagen, um auf der Eingabe des Nutzers basierende kontextbezogene und kohärente Antworten zu generieren (OpenAI, o. D.-a). Aufgrund des umfangreichen Datenkorpus, der während des Trainingsprozesses verwendet wurde und auch widersprüchliche sowie inkorrekte Informationen enthält, kann ChatGPT 4 mitunter überzeugende, jedoch unzutreffende Informationen produzieren, sogenannte Halluzinationen. Daher ist es ratsam, die von ChatGPT 4 generierten Antworten kritisch zu prüfen, insbesondere wenn eine hohe Zuverlässigkeit der Information erforderlich ist (OpenAI et al., 2024).

### 2.1.2 Large Language Models

Large Language Models (LLMs) sind Modelle künstlicher Intelligenz, die auf einem umfangreichen Textkorpus trainiert werden und mittels Deep-Learning-Techniken dazu in der Lage sind, natürliche Sprache zu verstehen sowie Texte und andere Inhalte, wie Programmcode, zu generieren (IBM, o. D.). LLMs bestehen aus vielschichtigen neuronalen Netzen mit mehreren Milliarden Parametern und basieren typischerweise auf der Transformer-Architektur, welche von Forschern bei Google im Jahr 2017 vorgestellt wurde (IBM, o. D.; Vaswani et al., 2017). Das in dieser Bachelorarbeit betrachtete Modell GPT-4 ist ein solches LLM, wobei neben diesem noch weitere bedeutende Modelle wie Googles Gemini 1.5, Metas Llama 3 und Anthropics Claude 3 existieren (OpenAI et al., 2024; Pichai & Hassabis, 2024; Meta, o. D.; Anthropic, 2024). Während des *Pre-Trainings* werden die Parameter des Modells mittels Optimierungstechniken angepasst, um das Verständnis allgemeiner Sprachstrukturen und die Erkennung von Wortbeziehungen zu ermöglichen (Vaswani et al., 2017). Dies ermöglicht es den LLMs, präzise Antworten zu geben und vielseitig angewandt zu werden, unter anderem zur Informationsabfrage, zur Generierung, Zusammenfassung und Übersetzung von Texten sowie zur Erstellung von Quellcode (IBM, o. D.). LLMs besitzen zudem die Fähigkeit, für Aufgabenstellungen verwendet zu werden, für die sie kaum oder gar keine Trainingsdaten zur Verfügung hatten (Kelbert et al., 2024).

### 2.1.3 LeetCode

LeetCode ist eine Plattform, die Programmieraufgaben in unterschiedlichen Themenbereichen und Schwierigkeitsgraden (leicht, mittel, schwer) anbietet und dabei insgesamt 27 verschiedene Programmiersprachen, einschließlich kompilierter und interpretierter Sprachen sowie Skript- und Datenbanksprachen, unterstützt (LeetCode, 2024; LeetCode, o. D.-c). Das Ziel der Plattform ist es, die Fähigkeiten der Nutzer zu erweitern und sie auf technische Interviews vorzubereiten (LeetCode, o. D.-b). Jede Aufgabe enthält eine detaillierte Beschreibung mit Beispielen und Einschränkungen sowie eine Codevorlage zur Implementierung. Mit dem integrierten Interpreter bzw. Compiler kann der Code direkt auf der Plattform ausgeführt werden, wobei entweder die spezifische Fehlermeldung oder eine Bestätigung des erfolgreichen Bestehens aller Tests ausgegeben wird. Zusätzlich erfasst die Plattform die Speicher- und Laufzeitwerte des akzeptierten Codes, vergleicht diese mit den Werten der Lösungen anderer Nutzer und erstellt daraus ein Ranking, das dem Nutzer eine Einschätzung im Vergleich zu anderen ermöglicht. Darüber hinaus bietet die Plattform die Möglichkeit, sich innerhalb der Community über Lösungen auszutauschen (LeetCode, o. D.-d).

## 2.2 Aktueller Forschungsstand

Der nachfolgende Abschnitt untersucht die neuesten Erkenntnisse und Fortschritte in der Erforschung der Code-Generierungsfähigkeiten von ChatGPT 4 über verschiedene Programmiersprachen hinweg. Im Zuge der Literaturrecherche wurde deutlich, dass sich die vorherrschende Forschung primär auf eine Leistungsanalyse von ChatGPT 4 in einer oder wenigen Programmiersprachen konzentriert. Hierbei wird die Leistung in einer spezifischen Sprache repräsentativ für die generelle Code-Generierungsfähigkeit von ChatGPT 4 verwendet, wodurch jedoch potenzielle Unterschiede in der Effizienz des Modells über verschiedene Sprachen hinweg vernachlässigt werden. Zudem wird häufig ein Vergleich zwischen ChatGPT 4 und anderen LLMs vorgenommen, wobei der Fokus weniger auf den spezifischen Programmiersprachen liegt als vielmehr auf einem Vergleich der Modelle untereinander. Für diese Evaluierung werden in der Regel Benchmarks wie *HumanEval*, *EvalPlus*, *CoderEval* und *DS-1000* herangezogen (Chen et al., 2021; Liu et al., 2023; Yu et al., 2024; Lai et al., 2023). Während der Durchführung der Literaturrecherche wurde zudem festgestellt, dass sich der Großteil der Studien auf die Vorgängerversion 3.5 von ChatGPT bezieht, anstatt auf die in dieser Arbeit untersuchte aktuelle Version 4.

Diese Tendenz unterstreicht die Herausforderung, spezifische Studien zu finden, die sich explizit mit der Code-Generierungsfähigkeit von ChatGPT 4 auseinandersetzen und dabei einen direkten Vergleich zwischen verschiedenen Programmiersprachen ziehen, was die Notwendigkeit für weiterführende Forschungen in diesem Bereich verdeutlicht. In Anbetracht der bestehenden Forschungslücke richtet sich das Augenmerk auf ausgewählte Studien, die den gegenwärtigen Stand der Forschung repräsentieren. Zunächst wird eine besonders relevante Untersuchung erörtert, die eine komparative Analyse der Code-Generierungsfähigkeiten des Vorgängermodells ChatGPT 3.5 in verschiedenen Programmiersprachen durchführt. Darauf folgen Studien, die die Fähigkeiten zur Code-Generierung von ChatGPT 4 in einer begrenzten Anzahl von Programmiersprachen evaluieren. Abschließend konzentriert sich die Betrachtung auf Forschungsarbeiten, die einen Leistungsvergleich zwischen ChatGPT 4 und anderen LLMs in einer einzelnen Programmiersprache vornehmen.

Ein wegweisendes Beispiel für diese Bachelorarbeit stellt die Studie von Buscemi aus dem Jahr 2023 dar, welche die Fähigkeiten von ChatGPT zur Code-Generierung in zehn Programmiersprachen untersucht. Zu den untersuchten Sprachen zählen C, C++, Go, Julia, JavaScript, Python, Perl, Ruby, R und Smalltalk, wobei C, C++, Go, JavaScript, Python und Ruby auch Gegenstand der vorliegenden Arbeit sind. In der betrachteten Studie kommt die vorherige Version 3.5 von ChatGPT zum Einsatz, wobei in dieser Arbeit die aktuelle Version 4 verwendet wird. Die Studie von Buscemi (2023) bewertet die Korrektheit, die benötigte Zeit und die Länge des generierten Codes. In der vorliegenden Arbeit werden die Länge des Codes und die Generierungszeit nicht berücksichtigt. Stattdessen erfolgt zusätzlich zur Code-Korrektheit eine Analyse der Laufzeit- und Speichereffizienz sowie der aufgetretenen Fehler.

Buscemi (2023) nutzte zur Durchführung der Studie einen Datensatz, der aus 40 Aufgaben aus den Bereichen Data Science, Algorithmen, Sicherheit und Spieleentwicklung bestand. Diese Aufgaben wurden von Universitätswebseiten und Coding-Plattformen entnommen. Im Gegensatz dazu verwendet diese Arbeit ausschließlich Herausforderungen der Plattform LeetCode und bezieht 188 allgemeine Programmierprobleme in die Analyse ein. Ein wesentlicher Kritikpunkt an der Studie von Buscemi (2023) besteht darin, dass nicht überprüft wurde, ob ChatGPT 3.5 möglicherweise auf Lösungen der gestellten Probleme trainiert wurde. In der vorliegenden Arbeit wurde hingegen darauf geachtet, diese Möglichkeit auszuschließen.

Bezüglich der Korrektheit des generierten Codes zeigte sich, dass ChatGPT 3.5 über alle Programmiersprachen hinweg in 45,8 % der Fälle korrekten Code erzeugen konnte. Dabei erzielte das Modell die höchste Erfolgsrate in der Programmiersprache Julia mit 81,5 %, während die niedrigste Lösungsrate mit 7,3 % in C++ verzeichnet wurde. Insgesamt stellte sich jedoch heraus, dass ein Großteil des von ChatGPT 3.5 generierten Codes trotz der vergleichsweise einfachen Aufgabenstellung überwiegend nicht ausführbar war und die Wahl der Programmiersprache das Verständnis der Aufgabenanforderungen beeinflusste.

In Bezug auf die Themengebiete zeigte ChatGPT 3.5 ebenfalls spezifische Stärken und Schwächen. Das Modell war in den Programmiersprachen C, C++ und Smalltalk besonders effektiv bei Sicherheitsaufgaben, während es bei algorithmenbasierten Aufgaben in Go, JavaScript, Perl, Python und Ruby die besten Ergebnisse erzielte. In den Programmiersprachen Julia und R wurden hohe Leistungen im Bereich Data Science festgestellt, während in der Spieleentwicklung über alle Programmiersprachen hinweg niedrigere Erfolgsquoten beobachtet wurden.

Die Antwortzeiten variierten ebenfalls erheblich. ChatGPT 3.5 benötigte für C im Durchschnitt 60 % mehr Zeit als für andere Programmiersprachen, während C++-Code in etwa der Hälfte der durchschnittlichen Zeit generiert wurde. Bei zusätzlicher Betrachtung der Code-Länge stellte Buscemi (2023) weiterhin keine Korrelation zwischen der Länge des Codes und der Antwortzeit fest. Es wurde jedoch eine größere Variabilität in der Code-Länge im Vergleich zur Antwortzeit beobachtet. Daraus schließt Buscemi (2023), dass der zeitliche Aufwand für das Verständnis der Aufgabenstellung geringer ist als für die Konzeption und das Schreiben des Codes.

Buscemi (2023) kam zu dem Schluss, dass die Effektivität von ChatGPT 3.5 bei der Code-Generierung stark von der verwendeten Programmiersprache abhängt. In dynamisch typisierten Sprachen hoher Abstraktionsebene erzielt ChatGPT 3.5 bessere Ergebnisse als in statisch typisierten Sprachen mit niedriger Abstraktionsebene. Das Modell erweist sich in populären Programmiersprachen als effektiver, da für diese umfangreichere Datensätze verfügbar sind, was ein umfassenderes Training ermöglicht.

Im folgenden Abschnitt werden zwei wissenschaftliche Studien untersucht, die die Fähigkeiten von ChatGPT 4 zur Code-Generierung in einer begrenzten Anzahl von Programmiersprachen evaluieren.

Während die Studie von Buscemi (2023) ein breites Spektrum von Programmiersprachen abdeckt, fokussiert sich die neuere Untersuchung von Bucaioni et al. (2024) auf die Programmiersprachen C++ und Java. Dabei wurde ChatGPT 4 auf 120 Programmierprobleme aus neun Kategorien und drei Schwierigkeitsgraden (einfach, mittel und schwer) der Plattform LeetCode getestet. Wie die vorliegende Bachelorarbeit nutzten Bucaioni et al. (2024) dieselbe Quelle für Programmierprobleme, die aktuelle Version 4 von ChatGPT, eine umfangreiche Anzahl an Programmierproblemen sowie die Metriken der Lösungsrate und der Speicher- und Laufzeiteffizienz. Im Gegensatz dazu fokussierten sich die Autoren jedoch auf den Vergleich zwischen C++ und Java und zogen dabei einen Vergleich zwischen ChatGPT 4 und menschlichen Programmierern, was in der vorliegenden Bachelorarbeit nicht vorgenommen wird.

Wie in dieser Arbeit wurde es ChatGPT 4 über drei Iterationen hinweg ermöglicht, die Lösung, basierend auf bereitgestelltem Feedback, zu verbessern. Dabei wurden die Lösungsraten, anders als in der vorliegenden Arbeit, nach Iterationen getrennt betrachtet. Für C++ wurde festgestellt, dass bei einfachen bis mittelschweren Aufgaben 96,67 % der Lösungen die Testsuite von LeetCode bereits in der ersten Iteration erfolgreich durchliefen, was sich im zweiten Durchgang auf 100 % steigerte. Bei schwierigen Aufgaben erreichte ChatGPT 4 eine Erfolgsquote von 76,67 %, die in der zweiten Iteration auf 93,33 % anstieg und in der dritten Iteration konstant blieb. Die spezifischen Ergebnisse für Java wurden von den Autoren nicht offengelegt, jedoch erwähnten sie ähnliche Beobachtungen hinsichtlich der Lösungsraten. Diese mangelnde Transparenz ist kritisch zu betrachten.

Bei der Laufzeiteffizienz erreichte ChatGPT 4 für C++-Code Perzentile von 58, 59 und 56 bei einfachen, mittelschweren und schweren Aufgaben. Dies bedeutet, dass der Anteil der menschlichen Lösungen, die eine längere Laufzeit aufwiesen und somit qualitativ unterlegen waren, im Vergleich zu den von ChatGPT 4 generierten Lösungen 58 %, 59 % beziehungsweise 56 % betrug. Für Java-Code erreichte das Modell bei einfachen Aufgaben fast das 100. Perzentil, schnitt jedoch bei mittelschweren und schweren Aufgaben deutlich schlechter ab.

Hinsichtlich der Speichereffizienz rangierte der C++-Code im 58., 47. und 56. Perzentil bei einfachen, mittelschweren beziehungsweise schweren Problemen. Die durchschnittlichen Werte des Java-Codes wurden textuell nicht erläutert, was eine bedeutende Lücke darstellt.

Bezüglich dieser Metriken stellten Bucaioni et al. (2024) zudem fest, dass der Versuch, den Code bezüglich Laufzeit und Speichereffizienz durch Hinzufügen eines Vermerks zum Prompt zu optimieren, paradoxerweise zu einer Verschlechterung der Speicher- und Laufzeitwerte führte, während das Hinzufügen dieser Information die Lösungsrate verbesserte, was Bucaioni et al. (2024) zu der Vermutung veranlasste, dass hier ein Kompromiss zwischen Lösungsrate und Speicher- sowie Laufzeiteffizienz besteht.

Bucaioni et al. (2024) zogen das Fazit, dass ChatGPT 4 bei Aufgaben von einfacher bis mittlerer Komplexität in Java und C++ überzeugende Lösungsraten erzielt, während bei anspruchsvolleren Aufgaben eine geringere Effektivität zu beobachten ist.

Sie empfehlen daher eine Verbesserung der Programmierfähigkeiten von ChatGPT, da dessen Leistungen bei komplexen Aufgaben noch begrenzt erscheinen. Obwohl dieser Schluss nachvollziehbar ist, sollte betont werden, dass eine Lösungsrate von 93,33 % bei schwierigen Aufgaben dennoch ein beachtliches Ergebnis darstellt. Eine mögliche Erklärung für diesen hohen Wert könnte sein, dass in der Studie auch Aufgaben berücksichtigt wurden, die vor dem letzten Trainingsstand von ChatGPT 4 auftraten (OpenAI, 2023b). Dies könnte die Aussagekraft der Ergebnisse verringern, da das Modell die Lösungen möglicherweise im Voraus kannte.

Obgleich ChatGPT 4 beachtliche Ergebnisse hinsichtlich Laufzeit- und Speichereffizienz erzielen konnte, kommen Bucaioni et al. (2024) zu dem Schluss, dass die Leistungen des Modells hinsichtlich Laufzeit- und Speichereffizienz im Vergleich zu menschlichen Programmierern noch nicht optimal sind. Es ist jedoch zu beachten, dass die Annahme, hinter den Lösungen stünden ausschließlich menschliche Programmierer, möglicherweise nicht zutrifft. LeetCode wird, wie in dieser Literaturrecherche gezeigt, zunehmend in wissenschaftlichen Untersuchungen zu LLMs verwendet, weshalb die Vergleichsdaten auch nichtmenschliche Lösungen umfassen. Abschließend legen die Autoren nahe, dass ChatGPT 4 in gegenwärtiger Form vorzugsweise als ein unterstützendes Werkzeug denn als vollständiger Ersatz angesehen werden sollte.

Nach ähnlichem Prinzip untersucht die Studie von Zhang et al. (2023) die Fähigkeiten der Modelle GPT-3.5 und GPT-4 zur Generierung von Python-Code, einer Programmiersprache, die in der vorliegenden Bachelorarbeit ebenfalls betrachtet wird. Die Untersuchung basiert auf der Bewertung von Lösungen zu Programmieraufgaben, sogenannten *Katas*, von der Plattform Codewars, welche ein ähnliches Konzept wie die in dieser Arbeit verwendete Plattform LeetCode verfolgt. Diese Aufgaben decken ein breites Spektrum an Schwierigkeitsgraden ab: von *8 kyu* für die einfachsten bis hin zu *1 kyu* für die anspruchsvollsten Aufgaben. Die Autoren klassifizieren die Niveaus acht bis fünf als leicht, vier und drei als mittelschwer sowie zwei und eins als schwer. In ähnlicher Weise werden in der vorliegenden Arbeit die Aufgaben nach den Schwierigkeitsgraden leicht, mittel und schwer eingeteilt.

Obwohl der Fokus von Zhang et al. (2023) auf dem Vergleich der beiden Modelle untereinander und mit menschlichen Programmierern liegt – ein Aspekt, der in dieser Arbeit nicht berücksichtigt wurde –, liefert die Studie wichtige Einblicke in die Code-Generierungsfähigkeiten von GPT-4 in Python. Es ist wichtig hervorzuheben, dass die Untersuchung auf nur 24 Probleme beschränkt ist. Zudem führen die Autoren die Lösung eines Problems auf *2 kyu*-Level durch GPT-4 auf ein vorheriges Training des Modells zurück, was darauf hindeutet, dass GPT-4 einige Lösungen potenziell bereits im Vorhinein kannte. Die Aussagekraft der Ergebnisse ist dahingehend eingeschränkt.

Die Ergebnisse der Studie zeigten, dass beide Modelle Aufgaben bis zu einem Schwierigkeitsgrad von *4 kyu* effektiv lösen konnten. Bei höheren Schwierigkeitsgraden ab *3 kyu* stießen beide Modelle jedoch an ihre Grenzen. GPT-4 war dabei zuverlässiger und konnte alle 15 Aufgaben von *8 kyu* bis *4 kyu* erfolgreich lösen, wobei es für vier Aufgaben auf

Feedback angewiesen war. Zudem gelang es GPT-4, ein Problem auf dem *2 kyu*-Niveau zu lösen. Obwohl GPT-3.5 die meisten Aufgaben bis zum *4 kyu*-Level bewältigen konnte, traten mehr Schwierigkeiten auf, wobei das Modell für fünf der gelösten Aufgaben auf Feedback angewiesen war und vier der 15 Probleme bis zum *4 kyu*-Level nicht lösen konnte.

Beide Modelle zeigten ein gutes Verständnis für die Aufgaben und ihre Anforderungen, auch wenn die endgültige Lösung nicht immer korrekt war, und konnten ihre Lösungsansätze durch gegebenes Feedback verbessern. Allerdings führte Feedback über fünf Runden hinaus zu keinen weiteren Verbesserungen. Kritisch zu betrachten ist, dass das Feedback keinem einheitlichen Schema folgte, sondern vergleichsweise informell und willkürlich gewählt wurde. Dies ist ein Unterschied zu der vorliegenden Arbeit, da hier die Bereitstellung von Feedback auf maximal drei Iterationen begrenzt wurde und es einem einheitlichen Schema folgt.

Zhang et al. (2023) schließen, dass GPT-4 das Niveau von Junior-Programmierern erreicht hat. Sie betonen zwar das Potenzial von GPT-4, merken jedoch an, dass das Modell derzeit noch nicht in der Lage ist, menschliche Programmierer vollständig zu ersetzen. Gründe hierfür sind unter anderem das Fehlen kreativen Denkvermögens und die Unfähigkeit zur Selbstvalidierung der erzeugten Lösungen.

Im letzten Abschnitt werden zwei weitere wissenschaftliche Arbeiten hervorgehoben, die einen Leistungsvergleich von ChatGPT 4 mit anderen LLMs in einer spezifischen Programmiersprache durchführen.

In einer 2023 veröffentlichten Studie von Bubeck et al. wurden verschiedene LLMs, darunter GPT-4, text-davinci-003 (ChatGPT 3.5), Codex (code-davinci-002) und CODEGEN-16B, hinsichtlich der Lösung der 164 Python-Programmierprobleme des HumanEval-Benchmarks sowie auf 100 LeetCode-Probleme der Sprache Python untersucht und verglichen. Zusätzlich wurden die Modelle GPT-3.5 und GPT-4 in realistischeren Anwendungsszenarien evaluiert. In der vorliegenden Bachelorarbeit wurde ein solcher Vergleich nicht durchgeführt, da der Fokus auf dem Vergleich der Code-Generierungsfähigkeiten von ChatGPT 4 in verschiedenen Programmiersprachen liegt.

Die Ergebnisse des ersten Benchmarks zeigen, dass GPT-4 mit einer Erfolgsquote von 82 % bei der Lösung der HumanEval-Probleme deutlich vor text-davinci-003 mit 65 %, Codex mit 39 % und CODEGEN-16B mit 30 % lag, was eine Überlegenheit des Modells zeigte. Die Autoren äußerten jedoch die Vermutung, dass GPT-4 möglicherweise speziell auf diese Art von Aufgaben trainiert worden sein könnte.

Um eine potenzielle Verzerrung auszuschließen, wurden zusätzlich 100 LeetCode-Probleme der Programmiersprache Python herangezogen, die nach dem 8. Oktober 2022 veröffentlicht wurden. Dieses Datum liegt nach dem Ende der Trainingsperiode von GPT-4 zum Zeitpunkt der Studienerstellung, was einen unvoreingenommenen Vergleich gewährleisten sollte. Ein ähnliches Vorgehen wurde auch in der vorliegenden Bachelorarbeit angewandt.

Für diesen zweiten Benchmark verwendeten Bubeck et al. (2023) die *pass@k*-Bewertungsmethode, bei der das Modell *k* Lösungen generierte und anschließend geprüft wurde, ob darunter eine korrekte Lösung vorlag. Diese Methode unterscheidet sich von der in der vorliegenden Arbeit verwendeten, insbesondere darin, dass der Code in den Versuchen der Studie von Bubeck et al. (2023) nicht durch Feedback verbessert wurde, während dies in der Bachelorarbeit der Fall ist. Darüber hinaus beschränkt sich die Bachelorarbeit auf drei Iterationen, während Bubeck et al. (2023) die Erfolgsrate sowohl nach einem als auch nach fünf Versuchen maßen.

Bei den einfachen Problemen erreichte GPT-4 eine Erfolgsquote von 68,2 % in einem und 86,4 % in fünf Versuchen. Für die mittelschweren Probleme lag die Erfolgsrate bei 40 % in einem Versuch und erhöhte sich auf 60 % nach fünf Versuchen. Bei den schweren Problemen erzielte GPT-4 eine Erfolgsrate von 10,7 % im ersten Versuch und 14,3 % nach fünf Versuchen, was in einer Gesamterfolgsrate von 38 % im ersten Versuch und 53 % in fünf Versuchen resultierte. Im Vergleich dazu erreichte text-davinci-003 eine Gesamterfolgsrate von 19 % im ersten Versuch und 36 % nach fünf Versuchen. Codex verzeichnete eine Gesamterfolgsrate von 13 % im ersten Versuch und 23 % nach fünf Versuchen. Menschliche Programmierer, spezifisch LeetCode-Nutzer, erreichten eine Rate von 38,2 %. Bubeck et al. (2023) folgern, dass GPT-4 den anderen Modellen in dieser Hinsicht überlegen ist und vergleichbare Ergebnisse wie menschliche Programmierer liefert. Es ist jedoch kritisch anzumerken, wie bereits bei der Studie von Bucaioni et al. (2024) angemerkt, dass es sich bei den LeetCode-Nutzern möglicherweise nicht ausschließlich um Menschen handelt.

Zusätzlich führten die Autoren einen dritten Benchmark durch, um die Fähigkeiten von GPT-4 im Vergleich zu GPT-3.5 bei realistischeren Programmierherausforderungen zu evaluieren. Dies stellt eine sinnvolle Erweiterung der Analyse dar, welche in der vorliegenden Bachelorarbeit nicht durchgeführt wurde, da dies über den vorgesehenen Umfang hinausgehen würde. Die untersuchten Aufgaben umfassen die Bereiche der Datenvisualisierung, LaTeX, Front-End-Entwicklung und Deep Learning.

Im Bereich der Datenvisualisierung extrahierten beide Modelle erfolgreich Daten aus LaTeX-Code, um Diagramme in Python zu erstellen. GPT-4 übertraf GPT-3.5 deutlich, indem es korrekte Diagramme erstellte und effektiv auf Anfragen zur weiteren Datenmanipulation reagierte, während GPT-3.5 bereits bei der Erstellung korrekter Diagramme scheiterte. In der Front-End- und Spieleentwicklung konnte GPT-4 funktionsfähigen Code für ein komplexes 3D-Spiel in HTML und JavaScript generieren, während GPT-3.5 sich weigerte, den angeforderten Code zu erstellen. Im Bereich Deep Learning sollten die Modelle ein benutzerdefiniertes Optimierungsmodul schreiben. Beide produzierten syntaktisch korrekten Code, aber nur GPT-4 erfüllte weitgehend die Instruktionen, während GPT-3.5 signifikante Fehler machte. Bei der Verarbeitung von LaTeX-Code konvertierte GPT-4 fehlerhaften Code erfolgreich in ausführbare LaTeX-Kommandos, während GPT-3.5 diese Aufgabe misslang.

Bubeck et al. (2023) kommen zu dem Schluss, dass GPT-4 im Vergleich zu den anderen untersuchten Modellen eine signifikante Überlegenheit aufweist, sowohl hinsichtlich des HumanEval-Datensatzes als auch bei der Lösung von 100 LeetCode-Problemen. Darüber hinaus zeigt GPT-4 gegenüber GPT-3.5 überlegene Fähigkeiten bei realistischeren Programmieraufgaben, was die erweiterte Anwendbarkeit des Modells in verschiedenen Programmierkontexten hervorhebt.

Die abschließende Arbeit dieser Literaturrecherche bildet die Studie von Coello et al. (2024). Ähnlich wie in der Studie von Bubeck et al. (2023) werden in dieser Untersuchung die Code-Generierungsfähigkeiten verschiedener LLMs, darunter GPT-4, GPT-3.5, Anthropics Claude, das auf GPT-4 basierende Bing sowie Googles Bard, untersucht. Der Fokus liegt dabei auf der Leistung dieser Modelle beim Lösen von 460 Problemen aus dem *Mostly Basic Python Problems* (MBPP)-Datensatz, der von Google-Forschern entwickelt wurde und sich an Programmieranfänger richtet. Dies stellt einen bedeutenden Unterschied zu dem in der vorliegenden Bachelorarbeit verwendeten Datensatz dar, da dieser auch Probleme höherer Schwierigkeitsgrade umfasst, die für Anfänger ungeeignet sind. Die Beschränkung auf einfache Programmierprobleme sollte beim Betrachten der Ergebnisse berücksichtigt werden. Des Weiteren untersuchten Coello et al. (2024) die Fähigkeit von GPT-4 und Bard, Feedback zu verstehen und ihre Code-Lösungen entsprechend zu verbessern.

Im ersten Benchmark zeigten sich deutliche Leistungsunterschiede zwischen den Modellen. Claude erreichte mit einer Lösungsrate von 71,43 % das schlechteste Ergebnis, gefolgt von Google Bard mit 76,16 %. Bing, basierend auf GPT-4, erzielte eine Lösungsrate von 81,96 %, während GPT-3.5 mit einem Wert von 83,18 % knapp darüber lag. An der Spitze stand GPT-4 mit 87,51 %. Eine bemerkenswerte Beobachtung war, dass nicht auf der GPT-Architektur basierende Modelle tendenziell längeren Code generierten, während GPT-basierte Modelle effizienteren, präziseren und kompakteren Code erzeugten.

Im zweiten Vergleich testeten Coello et al. (2024) die Fähigkeit von GPT-4 und Bard, aus manuell bereitgestelltem Feedback zu lernen und ihre Code-Lösungen entsprechend anzupassen. Es ist hierbei kritisch anzumerken, dass die Beschaffenheit des Feedbacks nicht offengelegt wurde. Während GPT-4 bei 16 nicht im ersten Anlauf gelösten Aufgaben 14 davon im zweiten Versuch mit Hilfe von Feedback korrigieren konnte, gelang dies Bard nur bei fünf von 16 Aufgaben.

Die Autoren kommen zu dem Schluss, dass GPT-4 im Vergleich zu anderen LLMs eine Überlegenheit bei der Lösung von Python-Aufgaben zeigt. Dabei weisen die GPT-Modelle die höchsten Kompetenzen auf und die Modelle Claude und Bard die niedrigsten. Zudem unterstreichen die Autoren die Effektivität von GPT-4 im Umgang mit Feedback und dessen potenzielle Nützlichkeit als Programmierassistent. Coello et al. (2024) betonen jedoch, dass GPT-4 und die anderen betrachteten LLMs die Notwendigkeit menschlicher Überwachung und Feedback nicht eliminieren und somit aktuell nicht als vollständiger Ersatz für menschliche Programmierer betrachtet werden können.

Die vorangegangenen Studien zeigen die Wichtigkeit einer vergleichenden Analyse der Code-Generierungsfähigkeiten von ChatGPT 4 über verschiedene Programmiersprachen hinweg. Eine solche Untersuchung ist essenziell, um die bestehende wissenschaftliche Forschung zu ergänzen und eine präzise Bewertung der aktuellen Fähigkeiten von ChatGPT 4 in der Code-Generierung zu ermöglichen, die wiederum die Basis für mögliche

Modellverbesserungen darstellen könnte. Dabei wird sichergestellt, dass ChatGPT 4 nicht vorab auf spezifische Lösungen trainiert wurde. Die Forschung berücksichtigt Aufgaben unterschiedlicher Komplexitätsstufen und verwendet einen zuverlässigen und umfangreichen Datensatz, um fundierte Schlussfolgerungen zu ermöglichen. Zudem erhält ChatGPT 4 die Möglichkeit, den Code durch Feedback zu optimieren, das nach einem einheitlichen Schema strukturiert ist. Im Zentrum der Analyse stehen die Lösungsraten in verschiedenen Programmiersprachen und eine Fehleranalyse, die darauf abzielt, spezifische Schwächen in den Programmiersprachen zu identifizieren. Die Untersuchung umfasst weiterhin eine Bewertung der Laufzeit- und Speichereffizienz im Vergleich zu Lösungen anderer Nutzer, um eine fundierte Beurteilung der Code-Qualität des von ChatGPT 4 generierten Codes zu ermöglichen.

# 3 Methodik

Das Ziel dieser Bachelorarbeit besteht darin, die Code-Generierungsfähigkeiten von ChatGPT 4 in 19 verschiedenen Programmiersprachen zu untersuchen. Durch einen systematischen Vergleich und eine detaillierte Analyse werden sowohl die Stärken als auch die Schwächen von ChatGPT 4 zwischen diesen Sprachen identifiziert. Zu diesem Zweck werden zwei Forschungsmethoden angewendet: eine Literaturrecherche und ein quantitatives Experiment. Dabei wird einem induktiven Forschungsansatz gefolgt, bei dem auf Basis der experimentell gewonnenen Erkenntnisse generalisierte Aussagen über die Leistungsfähigkeit von ChatGPT 4 in den betrachteten Programmiersprachen abgeleitet werden.

Um einen Überblick über den aktuellen Forschungsstand bezüglich der Code-Generierungsfähigkeit von ChatGPT 4 zu erlangen und die aktuellen Erkenntnisse mit den Ergebnissen dieser Arbeit vergleichen zu können, wurde eine Literaturrecherche unter Verwendung des PRISMA-Flussdiagramms vollzogen (Page et al., 2021). Die Resultate dieser Recherche sind in Abschnitt 2.2 dargestellt. Diese Methode wurde gewählt, da sie die Möglichkeit bietet, einen umfassenden Einblick in die aktuellen Forschungsbefunde zu gewähren und dabei einen konzeptuell fundierten Überblick über die zentralen Fakten und Daten zu vermitteln.

Zur spezifischen Beantwortung der Forschungsfragen wird zudem ein quantitatives Experiment durchgeführt. Dabei werden Daten zur Lösungsrate, zu aufgetretenen Fehlern sowie zur Laufzeit- und Speichereffizienz von Lösungen, die von ChatGPT 4 in den 19 Programmiersprachen generiert wurden, mit einem strukturierten Ansatz erhoben und anschließend einer deskriptiven statistischen Analyse unterzogen. Die verwendeten Programmierprobleme stammen von der Plattform LeetCode, die eine Vielzahl an Programmieraufgaben aus verschiedenen Themengebieten und den drei Schwierigkeitsgraden leicht, mittel und schwer anbietet (LeetCode, o. D.-c). Um ein möglichst vielfältiges und breites Spektrum an Programmierproblemen abzudecken, werden alle zum Zeitpunkt der Datenerhebung verfügbaren Probleme einbezogen, die nach dem letzten Wissensstand von ChatGPT 4 im April 2023 erstellt wurden (OpenAI, 2023b). Dieses Vorgehen gewährleistet, dass die vorhandene Datenbasis vollständig ausgeschöpft wird und ChatGPT 4 nicht vorab auf diese Probleme trainiert worden sein kann. Die von ChatGPT 4 generierten Code-Lösungen werden im Anschluss, abhängig von der verwendeten Programmiersprache, entweder mittels des auf LeetCode integrierten Compilers kompiliert oder durch den Interpreter ausgeführt und mehreren Testfällen unterzogen. Nach Abschluss der Tests werden sämtliche Daten gesammelt und in einem Git-Repository gespeichert. Sobald alle Aufgaben bearbeitet wurden, werden die gesammelten Informationen zusammengefasst und statistisch analysiert. Der gesamte Prozess wird dabei durch speziell für diese Bachelorarbeit entwickelte Python-Skripte unterstützt und automatisiert. Diese Methodik wurde gewählt, weil sie eine systematische und objektive Erfassung von Daten ermöglicht, die für die Beantwortung der Forschungsfragen essentiell ist. Durch die Durchführung eines quantitativen Experiments kann eine große Menge an Daten unter kontrollierten Bedingungen gesammelt werden, was die Zuverlässigkeit und Repräsentativität der Ergebnisse erhöht. Im Vergleich zu anderen

Methoden, wie qualitativen Interviews, bietet diese Methodik den Vorteil, quantifizierbare und objektiv vergleichbare Daten zu erheben, die eine solide Basis für die Beantwortung der Forschungsfragen bilden.

## 3.1 Literaturrecherche

Die Literaturrecherche wurde im März 2024 innerhalb einer Woche mithilfe von Google Scholar durchgeführt. Es wurden verschiedene Schlüsselbegriffe bezüglich ChatGPT und künstlicher Intelligenz mit relevanten Begriffen zur Code-Generierung verknüpft. Dabei wurden sowohl englische als auch deutsche Begriffe verwendet, um eine vollständige Erfassung zu ermöglichen. Die Suchformel hatte dabei folgende Struktur:

(„ChatGPT" OR „ChatGPT 4" OR „GPT" OR „GPT 4" OR „LLM" OR „AI" OR „KI" OR „Artificial Intelligence" OR „Künstliche Intelligenz") AND („Code Generation" OR „Code Generierung" OR „Coding" OR „Code" OR „Programming" OR „Programmieren")

Die Literaturrecherche bestand aus insgesamt drei Schritten. Zunächst wurde mithilfe der Suchformel eine Stichwortsuche durchgeführt, bei der 1164 Artikel identifiziert wurden. Anschließend wurden diese in eine Tabelle übertragen und 12 Duplikate entfernt. Danach wurde ein Screening-Prozess durchgeführt, bei dem die Zusammenfassungen der verbleibenden 1152 Artikel auf vorher definierte Ausschlusskriterien überprüft wurden. Diese Kriterien schlossen Artikel aus, die nicht in englischer oder deutscher Sprache verfasst wurden, keine empirischen Daten oder Analysen enthielten, keine thematische Relevanz hatten oder aufgrund von Zugangsbeschränkungen nicht zugänglich waren. In diesem Schritt wurden 1096 Arbeiten ausgeschlossen. Im dritten und letzten Schritt wurden die Volltexte von 56 Arbeiten gesichtet und ihre Eignung anhand von Einschlusskriterien bewertet. Die Einschlusskriterien bedingten, dass die Artikel einen quantitativen Benchmark von ChatGPT 4 bezüglich verschiedener Programmierprobleme unter Einsatz einer oder mehrerer Programmiersprachen enthalten, dass die Methodik eine angemessene Qualität und Aussagekraft aufweist und dass der Fokus auf der Code-Generierung liegt, nicht auf anderen Bereichen des Software-Engineerings. Zusätzlich wurden Arbeiten berücksichtigt, die einen relevanten Vergleich der Code-Generierungsfähigkeiten mehrerer Programmiersprachen mit ChatGPT 3.5 durchgeführt haben. Von diesen Arbeiten wurden 51 ausgeschlossen und fünf Berichte letztendlich einbezogen. Der gesamte Prozess ist in Abbildung 1 visualisiert.

**Abbildung 1**
*Suchprozess visualisiert als PRISMA-Flussdiagramm*

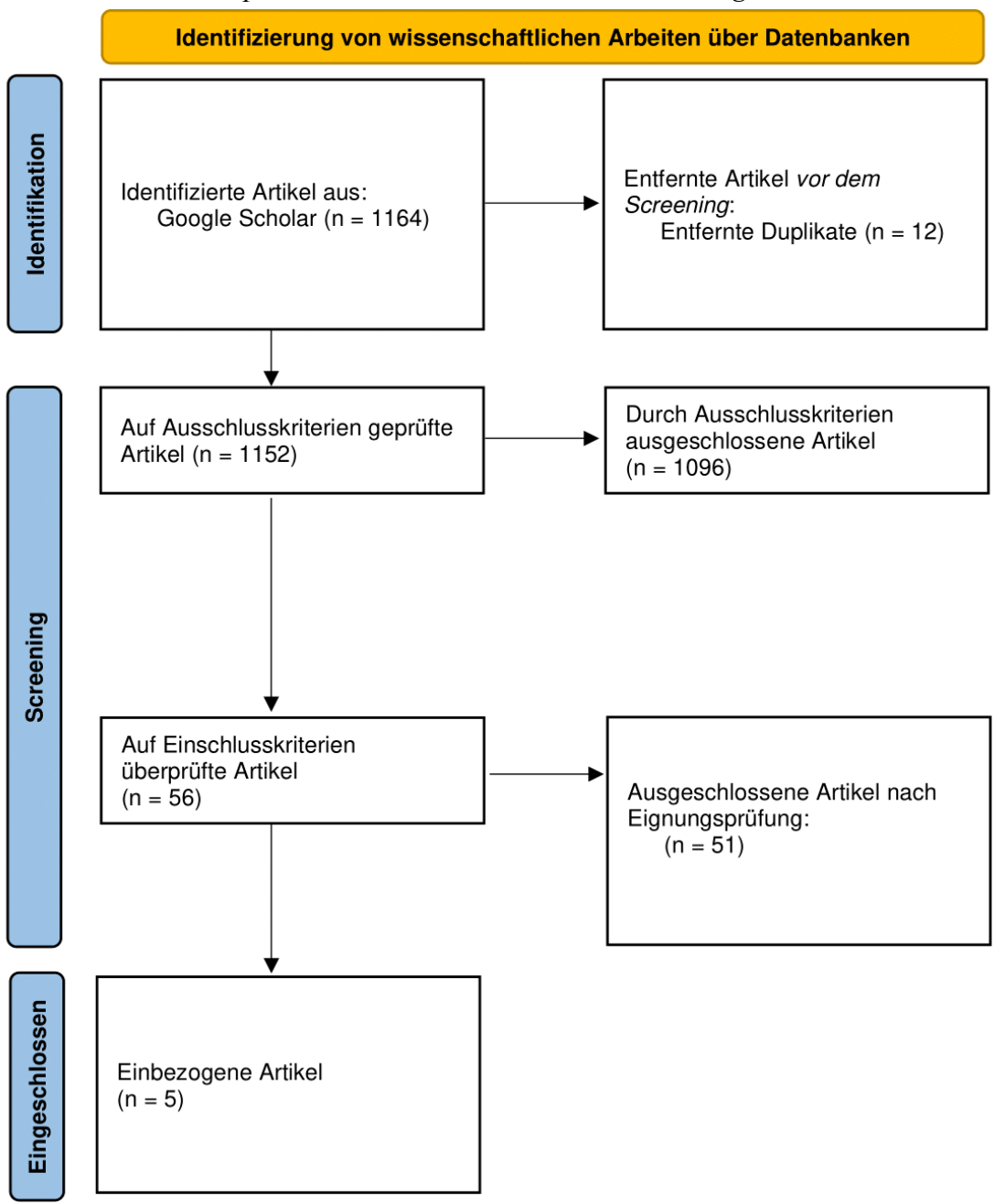

*Anmerkung.* In Anlehnung an Page et al., 2021

## 3.2 Experiment

Im folgenden Abschnitt wird das durchgeführte Experiment detailliert beschrieben. Zunächst wird ein Überblick über die ausgewählten Programmiersprachen gegeben, gefolgt von einer Darstellung der verwendeten Tools. Abschließend erfolgt eine detaillierte Beschreibung der Implementierung.

### 3.2.1 Überblick über die ausgewählten Programmiersprachen

Um die Code-Generierungsfähigkeiten von ChatGPT 4 in verschiedenen Programmiersprachen zu untersuchen, wurde eine Sammlung von insgesamt 19 Programmiersprachen erstellt. Dazu gehören Kotlin, Java, Rust, Scala, Go, C#, C++, PHP, Python (in den Versionen 2 und 3), TypeScript, Dart, JavaScript, Swift, Ruby, C, Racket, Elixir und Erlang. Diese Auswahl deckt ein breites Spektrum ab und umfasst gemäß aktuellen Umfragen und relevanten Rankings sowohl weit verbreitete als auch weniger häufig verwendete Programmiersprachen (*Stack Overflow Developer Survey 2023*, 2023; Cass, 2023; O'Grady, 2024). Sie sichert zudem eine umfassende Vielfalt, die die Untersuchung der Anpassungsfähigkeit von ChatGPT 4 an sprachspezifische Eigenheiten ermöglicht, wobei die Handhabung unterschiedlicher Syntax und Semantik jeder Sprache betrachtet werden kann. Die ausgewählten Sprachen decken zudem einen Erscheinungszeitraum von 1972 (C) bis 2014 (Swift) ab, wodurch sowohl moderne als auch historische Programmiersprachen Berücksichtigung finden (Ritchie, 1993; Goodwill & Matlock, 2015). Darüber hinaus ermöglicht die Betrachtung sowohl der aktuellen als auch einer älteren Python-Version eine Analyse der Fähigkeiten im Umgang mit Legacy-Code. Die Beschränkung auf diese 19 Sprachen gewährleistet einen Kompromiss zwischen Abdeckung und Durchführbarkeit der Untersuchung, ohne die Analyse zu überfrachten, und hält die Studie in einem angemessenen zeitlichen Rahmen. Die gewählte Zusammenstellung von Programmiersprachen verschiedener Paradigmen mit unterschiedlichen Eigenschaften stellt sicher, dass die Studie ein umfassendes Verständnis der Code-Generierungsfähigkeiten von ChatGPT 4 über ein weites Spektrum der Programmierlandschaft hinweg bietet. Bei allen Sprachen handelt es sich um Allzweck-Programmiersprachen, die nicht auf bestimmte Anwendungsbereiche beschränkt, jedoch auf einige spezialisiert sind. Detaillierte Kurzbeschreibungen der verwendeten Programmiersprachen sind in Tabelle 1 im Anhang zu finden.

### 3.2.2 Verwendete Tools

Zur Durchführung der Datenerhebung werden eine Reihe von Tools eingesetzt. Unter diesen Tools befinden sich ChatGPT 4, LeetCode, Git, GitHub sowie Python 3.10. Um Zugriff auf ChatGPT 4 zu erhalten, wurde ein kostenpflichtiges ChatGPT Plus-Abonnement (20 USD) abgeschlossen. Dies ermöglicht den Zugang zum Modell über das auf der OpenAI-Webseite bereitgestellte Webinterface (*ChatGPT*, o. D.). Die Entscheidung, LeetCode als Plattform zur Sammlung von Aufgaben und zur Überprüfung der Lösungen zu wählen, ist durch mehrere Faktoren begründet. Erstens ermöglicht die Plattform durch ihre umfangreiche Auswahl an Herausforderungen verschiedener Kategorien und Sprachen eine breite Datenerfassung, was die Validität der erhobenen Daten stärkt. Zweitens vereinfacht die integrierte Funktion zur Code-Auswertung signifikant die Analyse der Lösungsqualität und bietet Vergleichsdaten zu den Lösungen anderer Nutzer. An dritter Stelle ist die weitgehende Kostenfreiheit der meisten Aufgaben auf LeetCode zu nennen, die die Zugänglichkeit und Nachvollziehbarkeit der Forschungsergebnisse verbessert. Die Plattform zeichnet sich zudem durch eine umfangreiche Nutzerbasis aus, wodurch im Vergleich zu ähnlichen Plattformen wie HackerRank, CodeWars, SPOJ und CodeSignal bedeutende zusätzliche Vergleichsdaten zur Verfügung stehen. Dabei ist zu berücksichtigen, dass diese Einschätzung auf den geschätzten Besucherzahlen der Plattform Similarweb basiert. Diese Zahlen dienen als wertvolle Indikatoren, beanspruchen jedoch keine absolute Genauigkeit und sollten daher als Näherungswerte betrachtet werden (Similarweb, 2024a; Similarweb, 2024b; Similarweb, 2024c; Similarweb, 2024d; Similarweb, 2024e). Nicht zuletzt verdeutlicht die Nutzung von LeetCode durch OpenAI in eigenen Benchmarks, wie beispielsweise im technischen Bericht zu GPT-4, die Eignung der Plattform für wissenschaftliche Forschungszwecke (OpenAI et al., 2024).

Zur effektiven Erfassung und Verarbeitung der Daten wurden mehrere spezialisierte Python-Skripte erstellt. Das gesamte Projektmanagement erfolgte durch das Versionskontrollsystem Git, wobei der Code auf GitHub verfügbar gemacht wurde (Gilbert, 2024).[1] Durch diese strukturierte Herangehensweise und die Auswahl geeigneter Tools konnte das Experiment mit einer soliden Datenbasis und effizienten Arbeitsprozessen durchgeführt werden.

### 3.2.3 Implementierung

Um das Experiment zur Beantwortung der Forschungsfragen durchführen zu können, ist es zunächst notwendig, eine breite und diverse Sammlung von Programmieraufgaben zu erstellen. Zu diesem Zweck werden alle Aufgaben berücksichtigt, deren erste Lösung mindestens 14 Tage nach dem letzten Wissensstand von ChatGPT 4 eingereicht wurde. Dieser Wissensstand datiert zum Zeitpunkt der Erstellung dieser Bachelorarbeit auf den April 2023 (OpenAI, 2023b). Eine solche Vorgehensweise, die erste Lösung zur Identifikation des Veröffentlichungszeitraums heranzuziehen, wurde notwendigerweise gewählt, da eine direkte Abfrage des Veröffentlichungsdatums nicht möglich ist. Durch die Einhaltung einer zeitlichen Verzögerung wird zudem sichergestellt, dass den Nutzern genügend Zeit zur Verfügung stand, um Lösungen zu entwickeln. Diese Methodik gewährleistet einerseits die vollständige Nutzung der auf LeetCode verfügbaren Datenbasis und stellt andererseits sicher, dass ChatGPT 4 nicht auf Lösungen vorab trainiert wurde. Um diesen Prozess der Extraktion der Programmieraufgaben zu beschleunigen und zu automatisieren, wurde das Python-Skript *leetcode_question_scraper.py* entwickelt. Dieses Skript dient dazu, die Aufgaben abzufragen, zu filtern und anschließend zu speichern.

Das Skript beginnt damit, sämtliche Aufgaben vom LeetCode-Webserver zu extrahieren. Aus dieser Sammlung werden alle Aufgaben entfernt, die ausschließlich für Premium-Nutzer verfügbar sind, wodurch sichergestellt wird, dass die Forschungsergebnisse auf einer breiten und allgemein zugänglichen Datenbasis beruhen. Um potenzielle Verzerrungen in der Datenerhebung, die durch die Bildanalysefähigkeiten von ChatGPT 4 entstehen könnten, zu vermeiden, werden darüber hinaus Aufgaben ausgeschlossen, die Bilder enthalten. Zudem werden Probleme ausgeschlossen, deren erste Lösung vor dem 14. Mai hochgeladen wurde. Nach diesem Ausschlussprozess werden die Aufgabenstellungen und Code-Vorlagen der verbleibenden Probleme extrahiert, zu Prompts formuliert und systematisch in einem Git-Repository gespeichert. Ein Beispiel für einen solchen Prompt ist in Abbildung 2 dargestellt.

---

[1] https://github.com/DieserLaurenz/Leetcode-Gym

**Abbildung 2**

*Beispiel-Prompt zu einer Programmieraufgabe*

```
Solve the specified problem within the provided C# function template, without declaring the main function and
using only libraries from the standard library. Provide the code without explanation.

Template:

public class Solution {
    public int AccountBalanceAfterPurchase(int purchaseAmount) {

    }
}

Problem:

Initially, you have a bank account balance of 100 dollars.
You are given an integer purchaseAmount representing the amount you will spend on a purchase in dollars.
At the store where you will make the purchase, the purchase amount is rounded to the nearest multiple of 10.
In other words, you pay a non-negative amount, roundedAmount, such that roundedAmount is a multiple of 10
and abs(roundedAmount - purchaseAmount) is minimized.
If there is more than one nearest multiple of 10, the largest multiple is chosen.
Return an integer denoting your account balance after making a purchase worth purchaseAmount dollars from the store.
Note: 0 is considered to be a multiple of 10 in this problem.

Example 1:

Input: purchaseAmount = 9
Output: 90
Explanation: In this example, the nearest multiple of 10 to 9 is 10. Hence, your account balance becomes 100 - 10 = 90.

Example 2:

Input: purchaseAmount = 15
Output: 80
Explanation: In this example, there are two nearest multiples of 10 to 15: 10 and 20. So, the larger multiple, 20, is chosen.
Hence, your account balance becomes 100 - 20 = 80.

Constraints:

0 <= purchaseAmount <= 100
```

*Anmerkung.* Quelle: Eigene Darstellung, Textquelle: LeetCode, o. D.-a

Der Prompt wird auf Englisch verfasst, um mögliche Missverständnisse zu vermeiden, die durch Übersetzungen von ChatGPT 4 entstehen könnten. Dies berücksichtigt zudem, dass die Problembeschreibungen auf LeetCode ausschließlich auf Englisch verfügbar sind. In initialen Tests wurde beobachtet, dass ChatGPT 4 dazu neigt, die Main-Funktion in einigen Programmiersprachen zu deklarieren, was zu Fehlern seitens des LeetCode-Compilers beziehungsweise Interpreters führt. Da dies nicht dem Zweck der Datenerhebung entspricht, wird ein entsprechender Hinweis hinzugefügt. Des Weiteren wird ein Vermerk zur Beschränkung auf die Standardbibliothek hinzugefügt, da LeetCode nur diese zulässt. Die Erhebung der Programmierprobleme fand am 21.01.2024 statt. Es wurden 188 Probleme über 38 verschiedene Themen erhoben, darunter 55 einfache, 94 mittelschwere und 39 schwere Probleme.[2]

Nach der erfolgreichen Extraktion der Problemstellungen wird im zweiten Schritt das Skript *leetcode_gym.py* verwendet, um sämtliche gesammelten Aufgaben systematisch zu durchlaufen und die entsprechenden Prompts an ChatGPT 4 zu übermitteln. Dabei wird sichergestellt, dass für jeden Prompt eine eigene Konversation erstellt wird, um zu verhindern, dass ChatGPT 4 durch bereits in anderen Sprachen generierte Lösungen beeinflusst wird.

Sobald die Generierung des Codes durch ChatGPT 4 abgeschlossen ist, wird dieser zur Evaluierung an den LeetCode-Server gesendet. Dort wird der Code mehreren hundert Testfällen unterzogen, um dessen Korrektheit und Effizienz zu überprüfen. Werden alle Testfälle erfolgreich bestanden, speichert das Skript die vom Server zurückgegebenen Daten, einschließlich der Werte zur Laufzeit- und Speichereffizienz, zusammen mit dem Code und der Versuchsnummer. Anschließend wird zum nächsten Prompt übergegangen.

Sollte jedoch nicht jeder Testfall bestanden werden, wird neben dem Code und der Versuchsnummer zusätzlich die Fehlermeldung extrahiert und gespeichert. Basierend auf der Fehlermeldung wird daraufhin ein Fehlerprompt erstellt, der in derselben Konversation an ChatGPT 4 übermittelt wird. Dieser Schritt ermöglicht es dem Modell, den vorgeschlagenen Code auf Basis des erhaltenen Feedbacks zu überarbeiten und zu verbessern, was den realen Prozess der Code-Revision widerspiegelt. Der Fehlerprompt besitzt beispielhaft die in Abbildung 3 dargestellte Struktur.

**Abbildung 3**

*Beispiel eines Fehlerprompts*

> This solution is incorrect. Please provide a corrected code implementation, considering the outlined error details for enhancements. Provide the corrected code without explanation.
>
> Error type:
>
> Wrong Answer
>
> Error details:
>
> Solution produced a wrong answer. Test cases passed: 117/774
>
> Last executed input: Input 1: [1,2,6,4] Input 2: 3
>
> Received output: 2
>
> Expected output: 3

*Anmerkung.* Quelle: Eigene Darstellung

---

[2] Die Themen umfassen: Array, Backtracking, Binary Indexed Tree, Binary Search, Bit Manipulation, Bitmask, Brainteaser, Breadth-First Search, Combinatorics, Counting, Dynamic Programming, Enumeration, Graph, Greedy, Hash Function, Hash Table, Heap (Priority Queue), Math, Matrix, Memoization, Monotonic Queue, Monotonic Stack, Number Theory, Ordered Set, Prefix Sum, Queue, Rolling Hash, Segment Tree, Shortest Path, Simulation, Sliding Window, Sorting, Stack, String, String Matching, Trie, Two Pointers und Union Find.

Der Zyklus der Übermittlung des Codes, des Testens, der Bereitstellung des Feedbacks und der Verbesserung wird insgesamt höchstens dreimal durchlaufen. Diese Begrenzung auf drei Iterationen basiert auf vorläufigen Tests, die zeigten, dass ChatGPT 4 nach diesem Punkt keine signifikanten Verbesserungen mehr erzielen konnte. Zusätzlich hilft diese Begrenzung, den Zeitaufwand in einem angemessenen Rahmen zu halten, während dennoch ausreichend Gelegenheit für Verbesserungen gegeben wird.

Im nächsten Schritt wird das Skript *find_problematic_responses.py* eingesetzt, um potenziell fehlerhafte Antworten zu identifizieren und die Datenerhebung der entsprechenden Aufgaben wiederholen zu können. Hierbei wird überprüft, ob trotz bestandener Aufgabe ein weiterer Versuch mit einer folgenden Versuchsnummer vorliegt, ob mehrere Datensätze zu einem Versuch vorhanden sind und ob doppelter Code zwischen den Versuchen auftritt. Durch diesen zusätzlichen Schritt wird die Qualität und Zuverlässigkeit der gesammelten Daten weiter erhöht. Die Datenerhebung begann am 21.01.2024 und endete am 27.02.2024. Es wurden insgesamt 8345 Prompts an ChatGPT 4 gesendet.

Nach Abschluss der Datenerhebung werden die gesammelten Daten zusammengeführt und in einer CSV-Datei abgespeichert, um sie für die nachfolgende Analyse vorzubereiten. Die gespeicherten Daten umfassen Informationen zu einer Vielzahl von Werten. Dazu zählen der Name des Problems, die Akzeptanzrate, die verwendete Programmiersprache, der Schwierigkeitsgrad, die zugehörigen Themen, der Status der Problemlösung, angegeben als *True* oder *False*, sowie die Versuchsnummer, in der eine Lösung ermittelt wurde. Zusätzlich werden die aufgetretenen Fehlermeldungen, die Laufzeit in Millisekunden, der Speicherverbrauch in Megabyte sowie die Perzentile dieser im Vergleich zu anderen LeetCode-Nutzern erfasst. Diese strukturierte Datenaufbereitung ermöglicht eine effiziente und zielgerichtete Analyse der Leistungsindikatoren.

Im letzten Schritt wird das Skript *analyse_results.py* verwendet, um eine statistische Analyse der erhobenen Daten durchzuführen. Die Analyse wird unter Einsatz der Python-Module *Matplotlib* und *Pandas* vorgenommen. Dabei werden die Anzahl an gelösten Problemen über verschiedene Programmiersprachen ermittelt, die aufgetretenen Fehlerarten je Sprache erfasst sowie die durchschnittlichen Laufzeit- und Speicherwerte berechnet. Die Ergebnisse werden mittels Diagrammen visualisiert, um Muster und Auffälligkeiten in der Leistung des Modells zu identifizieren und darzustellen.

# 4 Ergebnisse

Im folgenden Abschnitt werden die Ergebnisse in Bezug auf die Forschungsfragen dargestellt. Zunächst werden die Problemlösungsraten von ChatGPT 4 in den 19 Programmiersprachen präsentiert, unterteilt in die Gesamtlösungsrate über alle Aufgaben hinweg sowie nach den Schwierigkeitsgraden einfach, mittel und schwer (Forschungsfrage F1). Danach werden die aufgetretenen Fehlerarten detailliert nach Programmiersprache aufgezeigt (Forschungsfrage F2). Abschließend erfolgt eine Darstellung der Codequalität, indem die Durchschnittswerte der Perzentile zur Laufzeit- und Speichereffizienz zwischen den verschiedenen Sprachen verglichen werden (Forschungsfrage F3).

## 4.1 F1: *Wie effektiv löst ChatGPT 4 allgemeine Programmierprobleme unterschiedlicher Schwierigkeitsgrade in 19 verschiedenen Programmiersprachen?*

Zunächst werden die erzielten Erfolgsquoten unter Berücksichtigung aller 188 Programmieraufgaben im Kontext der Gesamtlösungsraten detailliert aufgezeigt. Im Anschluss werden die Lösungsraten nach Schwierigkeitsgraden differenziert dargestellt. Dabei entfallen auf die Kategorie der einfachen Aufgaben 55 Aufgaben, auf die der mittelschweren Aufgaben 94 Aufgaben und auf die der schweren Aufgaben 39 Aufgaben. Die Einteilung der Programmierprobleme zu den Schwierigkeitsgraden wurde der Plattform LeetCode entnommen.

### 4.1.1 Gesamtlösungsraten

Um eine ganzheitliche Sicht auf die Lösungsrate von ChatGPT 4 in den untersuchten Programmiersprachen zu erhalten, wurde die Anzahl der gelösten Aufgaben pro Sprache über alle 188 Probleme hinweg aggregiert. Die entsprechenden Werte sind in Abbildung 4 veranschaulicht.

**Abbildung 4**

*Lösungsrate aller Probleme (188) nach Programmiersprache*

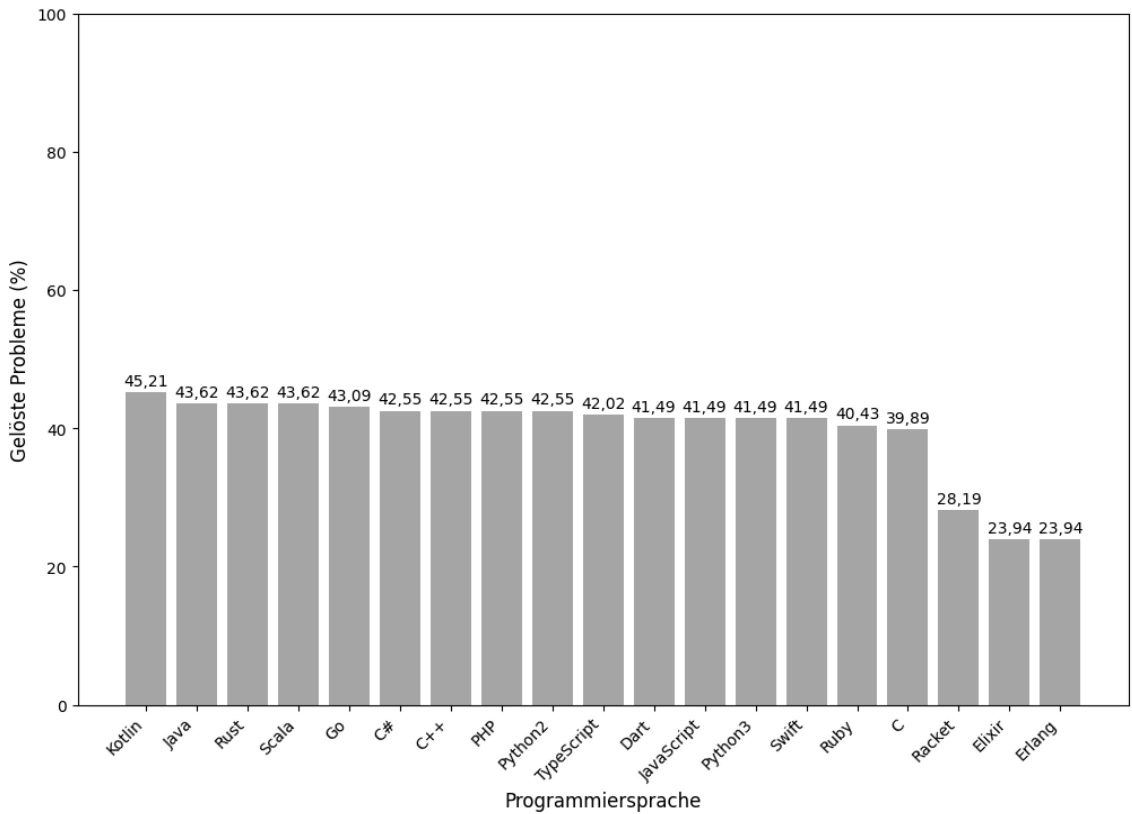

*Anmerkung.* Quelle: Eigene Darstellung

ChatGPT 4 erzielte über alle Sprachen und Aufgaben hinweg eine durchschnittliche Lösungsrate von 39,67 %, wobei das Modell in Kotlin mit 85 gelösten Aufgaben am erfolgreichsten war. Entgegen der Erwartung, dass Python aufgrund der Integration eines Code-Interpreters in ChatGPT 4 und der dadurch vermuteten umfassenderen Unterstützung die führende Sprache sein würde, erwies sich diese Annahme als unzutreffend (OpenAI, 2023a).

Es stellten sich zwei Cluster heraus, innerhalb derer vergleichbare Lösungsraten erzielt wurden. Der erste Cluster umfasste die ersten 16 Programmiersprachen. Dicht hinter Kotlin folgten hierbei die Sprachen Java, Rust und Scala mit jeweils 82 gelösten Aufgaben. Am unteren Ende befanden sich Ruby und C mit 76 beziehungsweise 75 gelösten Aufgaben. Wider Erwarten stellte sich heraus, dass ChatGPT 4 in Python 2 geringfügig effektiver war und zwei Aufgaben mehr bewältigen konnte als in der aktuellen Version Python 3.

Im zweiten Cluster zeigte sich im Vergleich zum ersten Cluster ein signifikanter Leistungsabfall. ChatGPT 4 hatte in den Programmiersprachen Racket, Erlang und Elixir größere Schwierigkeiten, Probleme zu lösen. In Racket erreichte ChatGPT 4 lediglich 53 gelöste Aufgaben, was den insgesamt drittletzten Platz bedeutete. Erlang und Elixir teilten sich den letzten Platz mit jeweils nur 45 gelösten Aufgaben.

Der beobachtete Leistungsabfall zwischen den Clustern war nicht unerwartet. Die Programmiersprachen Racket, Erlang und Elixir gehören zu denjenigen, die unter den betrachteten Sprachen am wenigsten verbreitet sind (*Stack Overflow Developer Survey 2023*, 2023; Cass, 2023; O'Grady, 2024). Dies legt nahe, dass für diese Programmiersprachen vermutlich weniger Trainingsdaten verfügbar waren, was letztlich zu niedrigeren Lösungsraten führte.

Im Kontext allgemeiner Programmierprobleme können C, C++, Rust und Go den Programmiersprachen mit niedrigem Abstraktionsniveau zugeordnet werden, während C#, Python 2, Python 3, Java, Kotlin, Scala, PHP, JavaScript, TypeScript, Ruby, Swift, Dart, Racket, Elixir und Erlang den Programmiersprachen mit höherem Abstraktionsniveau zugeordnet werden können. Bei dieser Zuordnung muss jedoch berücksichtigt werden, dass die Einteilung nach Abstraktionsniveaus je nach Betrachtungsweise und spezifischen Kriterien variieren kann. Es zeigte sich, dass ChatGPT 4 in Programmiersprachen mit niedrigem Abstraktionsniveau eine höhere Erfolgsrate von durchschnittlich 42,29 % gelöster Probleme erzielte, verglichen mit einer Erfolgsrate von 38,97 % in Programmiersprachen mit hohem Abstraktionsniveau.

Bei Betrachtung der Typisierung der Sprachen fiel auf, dass ChatGPT 4 in den statisch typisierten Sprachen C, C++, Rust, Go, C#, Java, Kotlin, Scala, TypeScript, Swift und Dart eine durchschnittliche Lösungsrate von 42,65 % aufwies und damit in diesen besser abschnitt als in den dynamisch typisierten Sprachen Python 2, Python 3, JavaScript, PHP, Ruby, Racket, Elixir und Erlang, wo eine Lösungsrate von lediglich 35,57 % erzielt wurde.

ChatGPT 4 erzielte somit tendenziell höhere Lösungsraten in statisch typisierten Programmiersprachen sowie in Programmiersprachen mit niedrigem Abstraktionsniveau im Vergleich zu dynamisch typisierten Programmiersprachen und solchen mit hohem Abstraktionsniveau.

## 4.1.2 Einfache Probleme

Abbildung 5 illustriert die Anzahl der gelösten Probleme von ChatGPT 4 nach Programmiersprache für die 55 Probleme, die dem einfachen Schwierigkeitsgrad zuzuordnen sind.

**Abbildung 5**

*Lösungsrate einfacher Probleme (55) nach Programmiersprache*

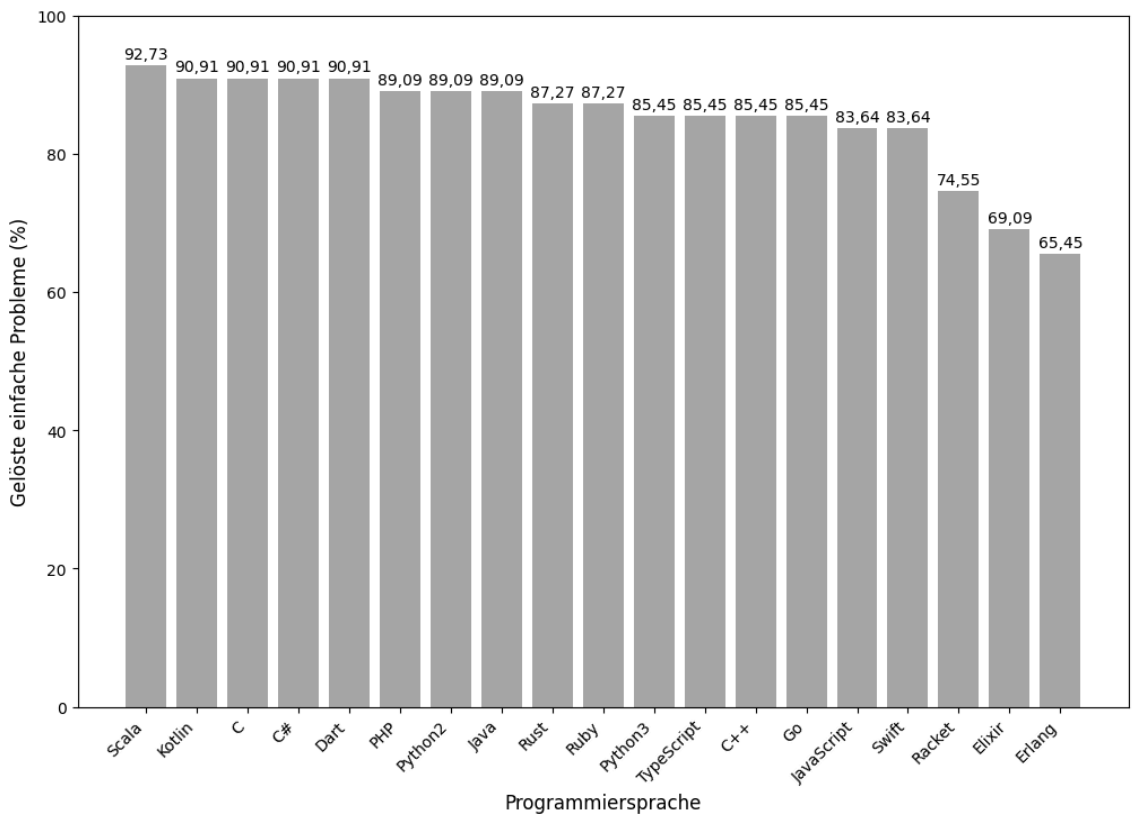

*Anmerkung.* Quelle: Eigene Darstellung

Die durchschnittliche Lösungsrate von ChatGPT 4 für einfache Programmierprobleme betrug 85,07 %. Generell erzielte ChatGPT 4 bei Aufgaben dieses Schwierigkeitsgrades hohe Erfolgsraten. Auch hier waren die zwei Cluster erkennbar, die bereits bei den Gesamtlösungsraten identifiziert wurden.

In der Programmiersprache Scala zeigte ChatGPT 4 mit 51 gelösten Aufgaben die beste Leistung, dicht gefolgt von Kotlin, C#, Dart und C, in denen jeweils 50 Aufgaben bewältigt wurden. Interessanterweise wies ChatGPT 4 in C, trotz einer relativ schwachen Gesamtlösungsrate, bei einfachen Problemen eine vergleichsweise hohe Leistung auf.

In den Sprachen Java, PHP und Python 2 bewältigte ChatGPT 4 jeweils 49 Aufgaben. Unmittelbar darauf folgten Rust und Ruby, in denen das Modell jeweils 48 Probleme erfolgreich löste. Weiterhin erzielte ChatGPT 4 in Go, C++, TypeScript und Python 3 jeweils eine Lösungsrate von 47 Aufgaben. Hierbei wurde der zuvor festgestellte Leistungsunterschied zwischen Python 2 und Python 3 erneut deutlich.

Am unteren Ende des Clusters befanden sich JavaScript und Swift, in denen ChatGPT 4 jeweils 46 Probleme lösen konnte. Die niedrigsten Lösungsraten verzeichnete das Modell, analog zu den Gesamtlösungsraten, in den weniger verbreiteten Programmiersprachen Racket, Elixir und Erlang. In Racket wurden 41 Probleme gelöst, was die drittniedrigste Leistung darstellte. Noch geringere Ergebnisse wurden in Elixir mit 38 gelösten Problemen erzielt, was die zweitschlechteste Leistung bedeutete. Die geringste Lösungsrate wurde in Erlang beobachtet, wo ChatGPT 4 lediglich 36 Probleme lösen konnte.

### 4.1.3 Mittelschwere Probleme

Die in Abbildung 6 dargestellten Werte zeigen die Lösungsraten von ChatGPT 4 für die 94 Probleme mittlerer Schwierigkeit, welche den größten Anteil der untersuchten Probleme ausmachen.

**Abbildung 6**

*Lösungsrate mittelschwerer Probleme (94) nach Programmiersprache*

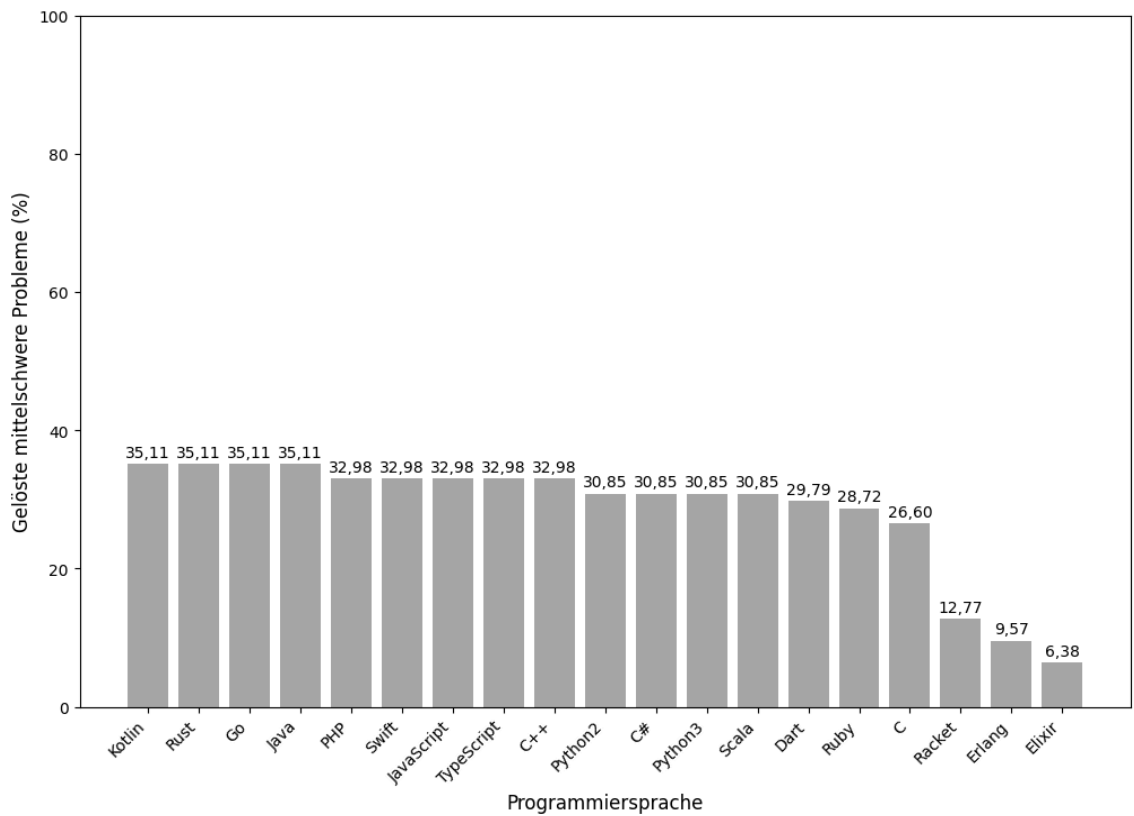

*Anmerkung.* Quelle: Eigene Darstellung

Die aggregierte Lösungsrate über alle betrachteten Programmiersprachen hinweg betrug 28,56 %, was im Vergleich zu den einfachen Problemen einen signifikanten Rückgang darstellte. Die beiden Cluster waren erneut deutlich zu erkennen.

ChatGPT 4 zeigte abermals eine hohe Lösungsrate in Kotlin sowie in den insgesamt starken Sprachen Rust, Go und Java, in denen jeweils die höchste Erfolgsquote von 33 gelösten

Problemen erzielt wurde. Darauf folgen PHP, Swift, JavaScript, TypeScript und C++ mit jeweils 31 gelösten Aufgaben. In Python, sowohl in den Versionen 2 als auch 3, sowie in C# und Scala, erreichte ChatGPT 4 jeweils 29 gelöste Probleme, womit sie knapp dahinter lagen.

Eine stärkere Abnahme der Lösungsraten war zwischen Dart, Ruby und C mit 28, 27 beziehungsweise 25 gelösten Problemen zu verzeichnen, wobei insbesondere die Schwierigkeiten von ChatGPT 4 mit C bei Problemen höheren Schwierigkeitsgrades deutlich wurden.

Schließlich waren die Herausforderungen, denen sich ChatGPT 4 in den Programmiersprachen Racket, Erlang und Elixir gegenübersah, erneut besonders auffällig, da hier lediglich 12, 9 beziehungsweise 6 Probleme gelöst werden konnten.

### 4.1.4 Schwere Probleme

In Abbildung 7 werden die Lösungsraten von ChatGPT 4 für die 39 schweren Probleme dargestellt. Die Abbildung illustriert klar die Herausforderungen, denen sich ChatGPT 4 bei der Bearbeitung dieser Aufgaben gegenübersah.

**Abbildung 7**

*Lösungsrate schwerer Probleme (39) nach Programmiersprache*

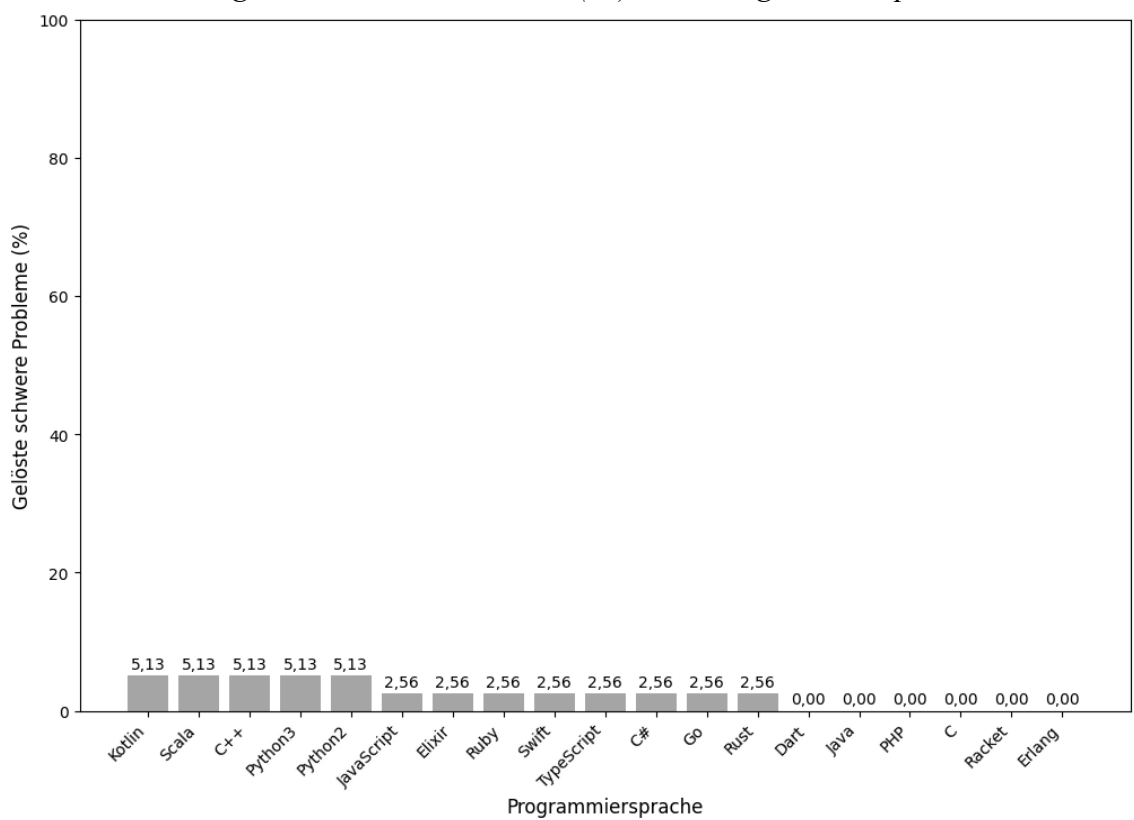

*Anmerkung.* Quelle: Eigene Darstellung

Die Lösungsrate über alle Sprachen hinweg betrug hierbei lediglich 2,43 %, was einen signifikanten Rückgang im Vergleich zu den mittelschweren Problemen und ein insgesamt schwaches Ergebnis darstellte.

In den Programmiersprachen Kotlin, Scala, C++, Python 3 und Python 2 bewältigte ChatGPT 4 jeweils zwei Probleme erfolgreich. Darüber hinaus gelang es ChatGPT 4 in den Sprachen JavaScript, Elixir, Ruby, Swift, TypeScript, C#, Go und Rust jeweils ein Problem zu lösen. Bemerkenswert war dabei die Lösung eines Problems in Elixir, da das Modell in dieser Sprache insgesamt eine vergleichsweise geringe Erfolgsrate aufwies. In den Sprachen Dart, Java, PHP, C, Racket und Erlang war ChatGPT 4 nicht in der Lage, ein einziges Problem erfolgreich zu bewältigen.

## 4.2 F2: *Welche Fehler treten in den von ChatGPT 4 in 19 verschiedenen Programmiersprachen generierten Lösungen auf?*

Um einen umfassenden Überblick über alle innerhalb der drei Iterationen aufgetretenen Fehler zu gewinnen, wurden die Fehlertypen akkumuliert und deren prozentuale Häufigkeiten in Abbildung 8 dargestellt. Die Fehlertypen unterteilen sich in die Kategorien *Wrong Answer*, *Time Limit Exceeded*, *Runtime Error*, *Memory Limit Exceeded* und *Compile Error*. Letzterer Fehlertyp konnte in den Sprachen JavaScript, Ruby, PHP, Python 2 und 3 sowie Dart nicht auftreten, da diese in der LeetCode-Umgebung interpretiert und nicht kompiliert wurden.

**Abbildung 8**
*Prozentualer Anteil der Fehlertypen nach Programmiersprache*

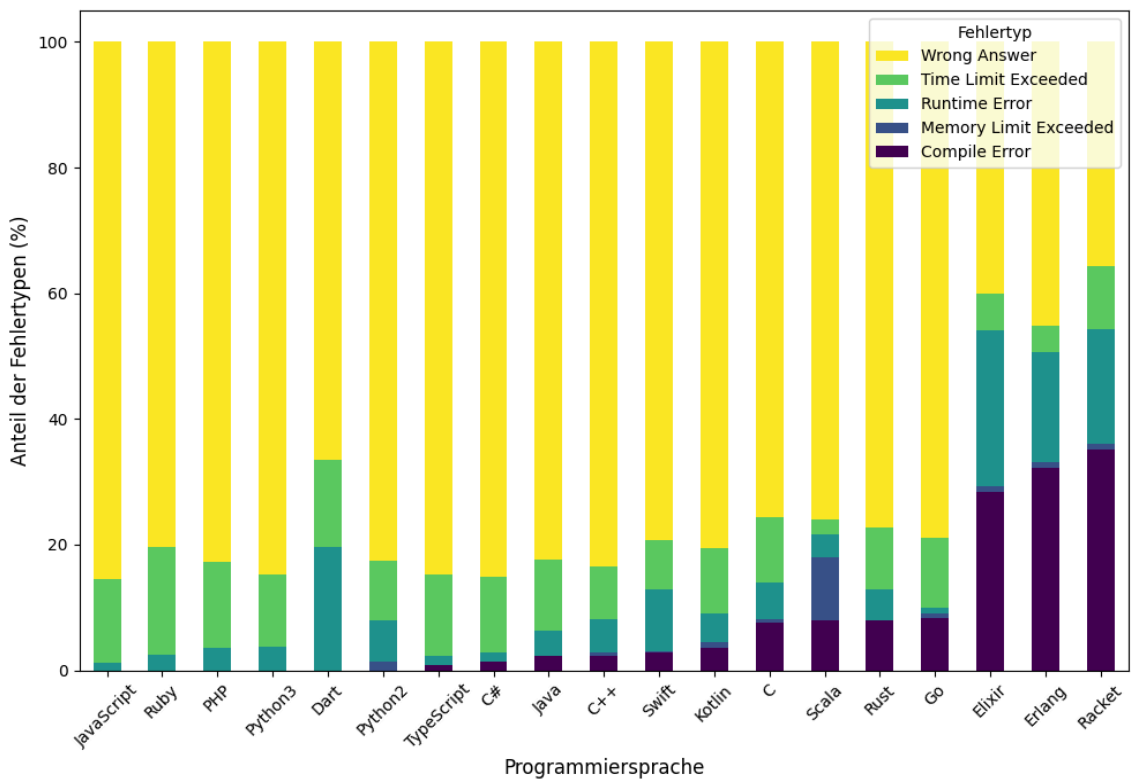

*Anmerkung.* Quelle: Eigene Darstellung

Es zeigte sich, dass ChatGPT 4 in 16 der 19 untersuchten Programmiersprachen bei fehlerhaften Antworten überwiegend syntaktisch korrekten und zur Laufzeit fehlerfreien Code generierte, der jedoch nicht alle Testfälle erfolgreich bestand, was zu dem Fehler *Wrong Answer* führte. Fehler in den restlichen drei Sprachen Elixir, Erlang und Racket waren hingegen hauptsächlich auf Compiler- und Laufzeitfehler zurückzuführen.

Der Fehler *Time Limit Exceeded* trat in den meisten Programmiersprachen proportional etwa gleich häufig auf. Insbesondere manifestierte sich dieser Fehler zwischen den Sprachen bei denselben Aufgaben, die einen Algorithmus mit hoher Zeitkomplexität erforderten und als mittelschwer oder schwer eingestuft wurden. Hierbei war ChatGPT 4 nicht in der Lage, einen ausreichend effizienten Algorithmus zu implementieren.

Der Fehler *Memory Limit Exceeded* trat nur sporadisch auf, wobei ChatGPT 4 diesem Fehler in Scala auffällig oft unterlag. Er trat zutage, wenn mehr Speicher verwendet wurde als vorgesehen, was auf eine dahingehend ineffiziente Codelösung hinwies.

Um einen tieferen Einblick in die Entstehung der Laufzeit- und Compilerfehler zu gewinnen, wurden die bereitgestellten spezifischen Fehlermeldungen analysiert und nach Fehlertyp kategorisiert, welche im Folgenden dargestellt werden. Es muss jedoch angemerkt werden, dass die vom Compiler und der Laufzeitumgebung ausgegebenen Fehlermeldungen nicht zwangsläufig die tatsächlichen Ursachen der Probleme widerspiegeln. Es ist möglich, dass ChatGPT 4 ursprünglich anderen Fehlern anheim fiel, die letztlich jedoch zu den gemeldeten Fehlern führten. Des Weiteren ist darauf hinzuweisen, dass aufgrund der umfassenden Komplexität einer derartigen Analyse nicht alle, sondern lediglich die bedeutendsten Fehler berücksichtigt werden konnten. Nichtsdestotrotz bieten die Ergebnisse einen interessanten Hinweis auf die spezifischen Schwächen von ChatGPT 4 in den Programmiersprachen und liefern Erkenntnisse darüber, welche Herausforderungen bei der Code-Generierung auftraten.

Zunächst werden die Programmiersprachen Elixir, Erlang und Racket separat betrachtet, da sie durch eine außergewöhnlich hohe Häufigkeit an Laufzeit- und Compilerfehlern auffielen und dementsprechend von den anderen Sprachen abzugrenzen sind. Im Anschluss daran werden die spezifischen Fehlertypen der übrigen Programmiersprachen untersucht. Zuerst werden die Compilerfehler und anschließend die Laufzeitfehler präsentiert, jeweils nach der Häufigkeit ihres Auftretens in den verschiedenen Programmiersprachen sortiert.

Bezüglich der Compilerfehler wurde in Elixir, Erlang und Racket eine hohe Anzahl an Syntaxfehlern identifiziert, welche in Form von inkorrekter Platzierung und Verwendung von Symbolen und Ausdrücken an ungeeigneten Stellen auftraten. Dies zeigte ein grundlegendes Missverständnis der strukturellen Prinzipien dieser Sprachen seitens ChatGPT 4. Spezifisch fehlten in Elixir und Racket häufig die schließenden Klammern, was dazu führte, dass offene Klammern nicht korrekt geschlossen wurden. In Racket wurden zudem gelegentlich fälschlicherweise eckige Klammern anstelle der erforderlichen runden Klammern zum Schließen verwendet. Des Weiteren wurde in Elixir unter anderem versucht, Schleifen mit dem Schlüsselwort *while* zu implementieren, obwohl Elixir While-Schleifen nativ nicht unterstützt. Ähnliche Missverständnisse zeigten sich in Erlang, wobei fälschlicherweise der in Elixir übliche, in Erlang jedoch nicht unterstützte Pipe-Operator ‚|>' verwendet wurde.

Abgesehen von den Syntaxfehlern trat in allen drei Sprachen der Versuch zutage, auf undefinierte Funktionen und Variablen zuzugreifen, was zu entsprechenden Fehlern führte. Dies ließ sich auf fehlende Definitionen, das Fehlen notwendiger Bibliotheken oder Module sowie auf das Versäumnis, eine benötigte Variable an die aufrufende Funktion weiterzugeben, zurückführen. In Elixir wurden zudem Variablen deklariert, jedoch später nicht verwendet, sowie in Erlang benutzerdefinierte Funktionen innerhalb von Guard-Expressions aufgerufen, was der Erlang-Compiler verbietet.

Nachdem die Compilerfehler identifiziert wurden, soll nun auf die Laufzeitfehler eingegangen werden. Bezüglich dieser traten in allen drei Programmiersprachen häufig Probleme im Zusammenhang mit Funktionsaufrufen auf, wo entweder eine falsche Anzahl an Argumenten oder der falsche Datentyp übergeben wurde. Beispielsweise bestand ein solcher Fehler darin, einer Funktion Nicht-Listentypen zu übergeben, obwohl Listen erforderlich waren.

Die restlichen Laufzeitfehler waren sprachspezifisch. Dabei wurde in Elixir häufig versucht, ähnlich wie in Python, auf Listen mittels eines Index zuzugreifen, was in Elixir jedoch nicht der empfohlenen Vorgehensweise entspricht. Ähnlich zu dem bereits erwähnten Compilerfehler wurden in Erlang während der Ausführung des Programms Funktionen aufgerufen, die nicht definiert waren. In Racket führte außerdem die Verwendung zu hoher Indizes für Listen ebenfalls zu Programmabstürzen.

Nachdem die spezifischen Fehler der Programmiersprachen Elixir, Erlang und Racket betrachtet wurden, folgt nun die Darstellung der Compiler- und Laufzeitfehler der verbleibenden Sprachen. Überraschenderweise traten auch in diesen Sprachen syntaktische und spezifische semantische Fehler auf, obwohl zuvor angenommen wurde, dass aufgrund der weiten Verbreitung und der umfassenden Dokumentation dieser Sprachen auftretende Fehler ausschließlich auf unzureichende Algorithmen zurückzuführen sein werden. Insbesondere stach Dart hier mit einem hohen Anteil an Laufzeitfehlern hervor. Im Folgenden werden alle Programmiersprachen genannt, in denen die Fehler aufgetreten sind. Aufgrund der Komplexität der Analyse werden jedoch nur die wesentlichen, detaillierten Fehler ausgewählter Programmiersprachen dargestellt, da eine vollständige Auflistung den Rahmen dieser Arbeit überschreiten würde.

Beginnend mit den Compilerfehlern zeigte sich, dass hierbei die häufigsten Fehler im Zusammenhang mit Datentypen auftraten. Diese Fehler traten speziell in den Sprachen Java, C#, TypeScript, Go, Rust, Scala, C, Kotlin, Swift und C++ zutage. In den Sprachen Java und C# trat ein spezifischer Fehler auf, bei dem einer Funktion eine Liste anstelle eines Arrays übergeben wurde und es so zu einer Verwechslung der Datentypen kam. Zudem zeigte sich in Java eine weitere Art dieses Missverständnisses, indem versucht wurde, auf Listen in der gleichen Weise zuzugreifen, wie es bei Arrays der Fall ist. Darüber hinaus kam es in Java zu Verwechslungen zwischen primitiven Typen und Referenztypen, indem versucht wurde, Methoden direkt auf den primitiven Typen *int* anzuwenden, obwohl solche Operationen in Java ausschließlich mit Referenztypen möglich sind. Weiterhin verwendete ChatGPT 4 in TypeScript fälschlicherweise die Bezeichnung *BigInt* zur Typisierung eines Arrays, obwohl der primitive Datentyp *bigint* gemeint war. Es ist anzumerken, dass *BigInt* lediglich die Konstruktorfunktion für den Datentyp *bigint* ist, sodass die Verwendung von *BigInt* als Typangabe nicht korrekt ist. Zudem gab es Versuche, den Plus-Operator zwischen den Typen *number* und *bigint* zu verwenden, was in TypeScript unzulässig ist. Schließlich nahm ChatGPT 4 in Go irrtümlich an, dass eine Funktion eine Liste von Integer-Werten zurückgegeben hatte, obwohl sie tatsächlich eine Map mit booleschen Werten lieferte. Dies führte zu dem Versuch, die Map wie eine Liste von Integer-Werten zu behandeln, was zu Typinkonsistenzen und folglich dazu führte, dass der Code nicht kompiliert werden konnte.

Die zweithäufigsten Compilerfehler waren Syntaxfehler, die in den Programmiersprachen Go, Swift, C++ und C auftraten. Beginnend mit Go versuchte ChatGPT 4 in einer Instanz eine ternäre Operation mit dem in anderen Sprachen üblichen Fragezeichen-Operator durchzuführen, obwohl dieser in Go nicht existiert. In Swift verwendete ChatGPT 4 irrtümlich einfache Anführungszeichen für String-Literale anstatt der korrekten doppelten Anführungszeichen. Zusätzlich ist zu nennen, dass ChatGPT 4 in C++ in einem Fall das Semikolon vergaß, um eine Anweisung formal korrekt abzuschließen. In diesem Fall ist es jedoch wahrscheinlich, dass dies ein Folgefehler war, der durch eine unzulässige Verwendung des Plus-Operators zwischen zwei Vektoren verursacht wurde, was den Compiler dazu brachte, auf ein Syntaxproblem hinzuweisen, das er nicht korrekt interpretieren konnte.

Anknüpfend lässt sich feststellen, dass der dritthäufigste Fehler aus der Deklaration der Main-Funktion resultierte, obwohl dies im Prompt ausdrücklich untersagt wurde. Dieser Fehler trat in den Programmiersprachen Go und C auf. Dies stellt ein Problem dar, da in der LeetCode-Umgebung ausschließlich die in der Code-Vorlage angegebene Funktion implementiert werden darf und die Main-Funktion nicht deklariert werden sollte.

Sämtliche verbleibenden Compilerfehler traten jeweils nur in einer Sprache auf. Einleitend mit Go importierte ChatGPT 4 hier in einigen Fällen die *Math*-Bibliothek, verwendete sie jedoch letztlich nicht, was der Go-Compiler als Fehler erkannte. In Rust unterliefen ChatGPT 4 bezüglich Trait-Bounds Fehler, indem die erforderlichen Traits *Hash* und *Eq* für das Hashing und den Vergleich der Schlüssel in *HashMap*-Strukturen nicht implementiert wurden. Dies führte bei einem Versuch, benutzerdefinierte Datentypen als Schlüssel zu verwenden, zu einem Kompilierungsfehler. In Swift wurden spezifische Fehler aufgrund von Mutabilitätskonflikten festgestellt, die durch den Versuch entstanden, unveränderliche Konstanten zu modifizieren. Zudem traten in Swift Fehler bei der Handhabung von Optionals auf, insbesondere durch ihre Verwendung ohne angemessenes Unwrapping. Abschließend kam es in Scala zu Fehlern durch fehlende Definitionen, da Code außerhalb von Klassen- oder Objektdefinitionen platziert wurde, was in Scala unzulässig ist.

Nachdem die Compilerfehler behandelt wurden, soll nun abschließend auf die Laufzeitfehler eingegangen werden. In diesem Kontext erwies sich der *Index Out of Range*-Fehler als besonders häufig. Dieser trat in den Sprachen Go, Scala, C, Kotlin, Swift, C++, Java, C# sowie Python 2 und 3 auf und resultierte aus dem Versuch, auf ein Array-Element zuzugreifen, dessen Index außerhalb der definierten Grenzen liegt.

Am zweithäufigsten traten erneut spezifische Fehler im Umgang mit Datentypen auf, die jedoch in diesem Fall zur Laufzeit erkannt wurden. Betroffen waren die Programmiersprachen Python 2 und 3, Dart, Ruby, JavaScript und TypeScript. Angefangen mit Python 2 versuchte ChatGPT 4 hier, die Methode *isdigit* auf einen Integer anzuwenden, obwohl diese Methode ausschließlich für Strings definiert ist. In Dart wurde ein Wert vom Typ *num* einer als *int* deklarierten Variablen zugewiesen, da versäumt wurde, das Ergebnis einer Division korrekt zu konvertieren. Darüber hinaus unterlief ChatGPT 4 ein spezifischer Fehler in Ruby, wo ein Boolean-Wert fälschlicherweise als Index für den Zugriff auf ein Array-Element verwendet wurde. Abschließend ist ein Fehler in JavaScript zu nennen, wo ein *bigint* ohne die erforderliche Konvertierung mit anderen Datentypen in einer Operation kombiniert wurde.

Der dritthäufigste Laufzeitfehler, den ChatGPT 4 verursachte, war die Division durch Null. Dieser Fehler trat in den Programmiersprachen Kotlin, C++, C#, Python 3, PHP, Dart und Ruby auf.

An vierter Stelle standen Fehler im Zusammenhang mit der Überschreitung von Serialisierungslimits, die auftraten, wenn versucht wurde, Werte zu serialisieren, die außerhalb der akzeptierten Bereiche lagen. Dieser Fehler kam in den vier Programmiersprachen Go, Kotlin, Swift und Scala vor.

In der gleichen Anzahl von Sprachen wurden zusätzlich Fehler bezüglich des Zugriffs auf undefinierte Methoden und Klassen beobachtet. Spezifisch handelte es sich hierbei um die Sprachen Dart, Ruby, JavaScript und PHP. Zum einen gab es in Dart den Versuch, die nicht existierende *Math*-Klasse zu verwenden, wobei eigentlich das in Kleinbuchstaben geschriebene *math*-Modul gemeint war. Zum anderen traten in Ruby und JavaScript Zugriffe auf undefinierte Methoden zutage, wie etwa der Aufruf der *bit_count*-Methode, die in Ruby nativ nicht existiert. Des Weiteren gab es den Versuch, die aus der GNU Compiler Collection (GCC) stammende Methode *__builtin_popcount* in JavaScript zu verwenden (*Other Builtins (Using The GNU Compiler Collection (GCC))*, o. D.).

Ein weiterer methodenbezogener Fehler, der in vier verschiedenen Programmiersprachen festgestellt wurde, bestand darin, dass Methoden mit einer inkorrekten Anzahl von Argumenten aufgerufen wurden. Dieser Fehler trat in den Programmiersprachen PHP, Ruby sowie Python 2 und 3 auf.

In den drei Sprachen Scala, Kotlin und Java wurden zusätzlich *Null Pointer Exceptions* registriert, die durch den Zugriff auf eine Objektreferenz, die den Wert *null* besitzt, verursacht wurden. In ähnlicher Weise traten in Swift Probleme mit fehlerhaften Dereferenzierungen von Pointern auf.

Ebenfalls in drei Sprachen wurden in Python 2 und 3, sowie in PHP *Value Errors* festgestellt. Spezifisch sind hier der Aufruf der *max*-Funktion auf einer leeren Sequenz, die Verwendung eines negativen Shift-Counts in Python 2 sowie die Nutzung negativer Werte für eine Anzahl an Wiederholungen in PHP zu nennen.

Auch wurden, ähnlich wie bei den Compilerfehlern, Syntaxfehler in Python 2, Dart und PHP identifiziert, die jedoch erst zur Laufzeit registriert wurden. Beispielsweise wurde in Python 2 ein fehlerhaftes doppeltes Leerzeichen zwischen Schlüsselwörtern festgestellt. In Dart fehlten Klammern um eine explizite Typumwandlung, was zu einer fehlerhaften Interpretation der Ausdrücke führte. Schließlich wurde in PHP eine nicht ausreichende Anzahl an Klammern verwendet, um eine formal korrekte Syntax einzuhalten.

In den drei C-Sprachen C, C++ und C# sind zusätzlich *Overflow*-Fehler aufgetreten. Diese Fehler manifestierten sich in C und C++ beispielsweise, als das Produkt zweier Zahlen den maximal zulässigen Wert für eine Integer-Variable überschritt. In C# hingegen trat dieser Fehler auf, wenn der Versuch unternommen wurde, einen String in einen *Int64*-Wert zu konvertieren, der die definierten Grenzen dieses Datentyps überstieg.

Zum Schluss ist festzustellen, dass einige Fehler nur in einzelnen Sprachen vorzufinden waren. So führten in Swift *Illegal Instructions* zu zahlreichen Programmabstürzen, wobei die genauen Ursachen nicht weiter spezifiziert wurden. In Python 2 wurde ein *Memory Error* identifiziert, der auftrat, wenn das Speichermanagement des Programms fehlerhaft war. In Python 3 traten *Key Errors* zutage, wenn versucht wurde, auf nicht vorhandene Schlüssel in Wörterbüchern zuzugreifen oder diese zu modifizieren. Abschließend wurden in PHP Deklarationen der Main-Funktion identifiziert, die ähnlich zu den bereits bei den Compilerfehlern erwähnten Fehlern sind, aber zur Laufzeit erkannt wurden.

## 4.3 F3: *Wie laufzeit- und speichereffizient sind die von ChatGPT 4 in 19 verschiedenen Programmiersprachen generierten Lösungen?*

Zur Bewertung der Qualität des von ChatGPT 4 generierten Codes wurden die durchschnittlichen Perzentile für die Laufzeit- und Speichereffizienz im Vergleich zu Lösungen anderer LeetCode-Nutzer über alle Problemstellungen und Programmiersprachen hinweg ermittelt. Die Perzentile geben an, wie eine bestimmte Lösung im Vergleich zu anderen Lösungen abschneidet. Ein Perzentil von 60 bedeutet beispielsweise, dass der Wert der betrachteten Lösung, sei es eine kürzere Laufzeit oder ein geringerer Speicherverbrauch, niedriger ist als der Wert von 60 % der Lösungen anderer Nutzer. Es wird zunächst auf die Werte bezüglich der Laufzeiteffizienz und anschließend auf die Werte der Speichereffizienz eingegangen.

## 4.3.1 Laufzeiteffizienz

Die in Abbildung 9 dargestellten Werte illustrieren die durchschnittlichen Laufzeitperzentile unter Berücksichtigung aller Programmierprobleme, gegliedert nach Programmiersprache.

**Abbildung 9**

*Durchschnittliches Laufzeitperzentil aller Probleme (188) nach Programmiersprache*

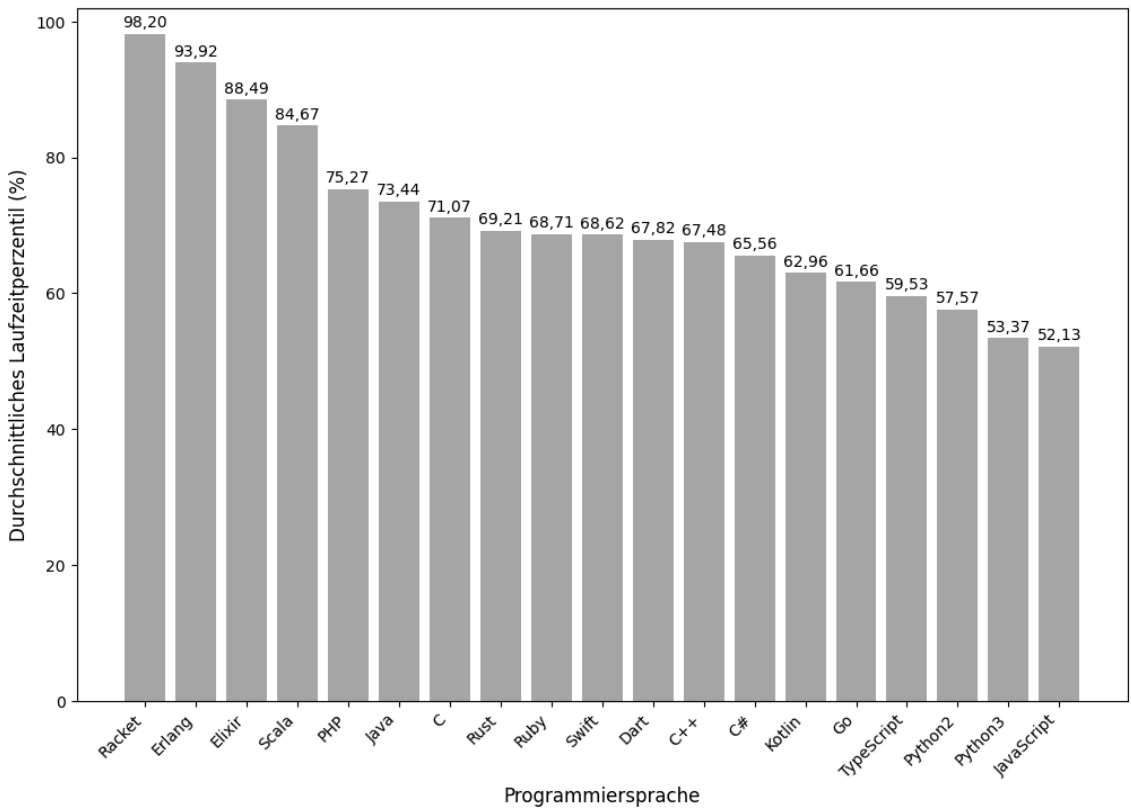

*Anmerkung.* Quelle: Eigene Darstellung

Es zeigte sich, dass ChatGPT 4 im Durchschnitt in allen untersuchten Programmiersprachen bessere Ergebnisse bezüglich der Laufzeiteffizienz erzielte als der durchschnittliche LeetCode-Nutzer. Insbesondere wies das Modell eine signifikante Stärke in den Sprachen Racket, Erlang und Elixir auf, obwohl in diesen die geringsten Probleme gelöst wurden. In Racket wurde im Durchschnitt sogar nahezu das 100. Perzentil erreicht, was bedeutet, dass in nahezu allen Programmierproblemen das beste Ergebnis im Vergleich zu allen anderen LeetCode-Nutzern erzielt wurde. Ähnliches galt für die Sprachen Erlang und Elixir, wobei hier ebenfalls in einigen, jedoch weniger Aufgaben das 100. Perzentil erreicht wurde.

Es ist hierbei wichtig anzumerken, dass die Ergebnisse möglicherweise durch die insgesamt geringe Anzahl an gelösten Aufgaben in diesen spezifischen Programmiersprachen verzerrt wurden. Darüber hinaus waren Lösungen von anderen LeetCode-Nutzern in diesen Sprachen generell selten und viele Probleme relativ neu, was die Vergleichsdatenmenge weiter reduzierte und die Repräsentativität der Daten einschränkte. Folglich kann nicht davon ausgegangen werden, dass ChatGPT 4 laufzeiteffizienten Code in diesen Sprachen generierte.

Weiterhin wies ChatGPT 4 signifikante Stärken in den Programmiersprachen Scala und PHP auf und erzielte zudem sehr solide Leistungen in Java und C, bei denen das Modell lediglich geringfügig hinter den besten 25 % aller Lösungen zurückblieb.

Zusätzlich erreichte ChatGPT 4 konsistente Ergebnisse in den Sprachen Rust, Ruby, Swift, Dart und C++, die alle im Bereich des 67. bis 70. Perzentils lagen. Für C#, Kotlin und Go zeigten die Ergebnisse einen deutlichen Rückgang, während TypeScript, Python 2, Python 3 und JavaScript die unteren Ränge einnahmen, welche dennoch geringfügig über den Ergebnissen des durchschnittlichen LeetCode-Nutzers lagen. Hierbei ist der nicht unerhebliche Leistungsunterschied zwischen TypeScript und JavaScript zu bemerken, ebenso wie die Tatsache, dass ChatGPT 4 in Python erneut, wenn auch geringfügig, bessere Ergebnisse in der älteren Version 2 als in der aktuellen Version 3 erzielte.

In Bezug auf das Abstraktionsniveau zeigte sich, dass ChatGPT 4 bei Sprachen mit höherem Abstraktionsniveau das 71,35. Perzentil erreichte, während das Modell bei Sprachen mit niedrigerem Abstraktionsniveau ein Perzentil von 67,35 erzielte. Die hohen Werte der Programmiersprachen Elixir, Erlang und Racket schienen diese Tendenz erheblich zu beeinflussen, weshalb es sinnvoll war, sie aufgrund ihrer verzerrenden Eigenschaften auszuschließen, um ein präziseres Bild zu erhalten. Bei Umsetzung dieser Maßnahme ergab sich ein verändertes Gesamtbild. In diesem Fall erreichte ChatGPT 4 in Sprachen mit höherem Abstraktionsniveau lediglich das 65,80. Perzentil, was somit im Gegensatz zu den zuvor festgestellten Werten unter dem Wert für Sprachen mit niedrigerem Abstraktionsniveau lag, wo erneut ein Perzentil von 67,35 erzielt wurde. ChatGPT 4 generierte folglich tendenziell laufzeiteffizienteren Code in Sprachen mit niedrigerem Abstraktionsniveau.

Bei Kategorisierung der Programmiersprachen nach ihrer Typisierung zeigte sich zunächst eine klare Bevorzugung dynamisch typisierter Sprachen mit einem Perzentil von 73,46 im Vergleich zu statisch typisierten Sprachen, wo ein Wert von 68,37 festgestellt wurde. Eine erneute Betrachtung der Daten, bei der die potenziell verzerrenden Werte von Elixir, Erlang und Racket ausgeschlossen wurden, führte zu einem veränderten Ergebnis. Ohne Berücksichtigung dieser Werte erreichte ChatGPT 4 in dynamisch typisierten Sprachen ein signifikant niedrigeres Ergebnis im 61,41. Perzentil. Im Vergleich lagen diese Werte, anders als zuvor, diesmal unter denen der statisch typisierten Sprachen, bei denen erneut ein Perzentil von 68,37 festgestellt wurde. Das Modell zeigte somit eine Präferenz für statisch typisierte Programmiersprachen in Bezug auf die Laufzeiteffizienz.

## 4.3.2 Speichereffizienz

In Abbildung 10 sind die durchschnittlichen Speicherperzentile, aufgeschlüsselt nach Programmiersprachen und unter Berücksichtigung aller Programmierprobleme, dargestellt.

**Abbildung 10**

*Durchschnittliches Speicherperzentil aller Probleme (188) nach Programmiersprache*

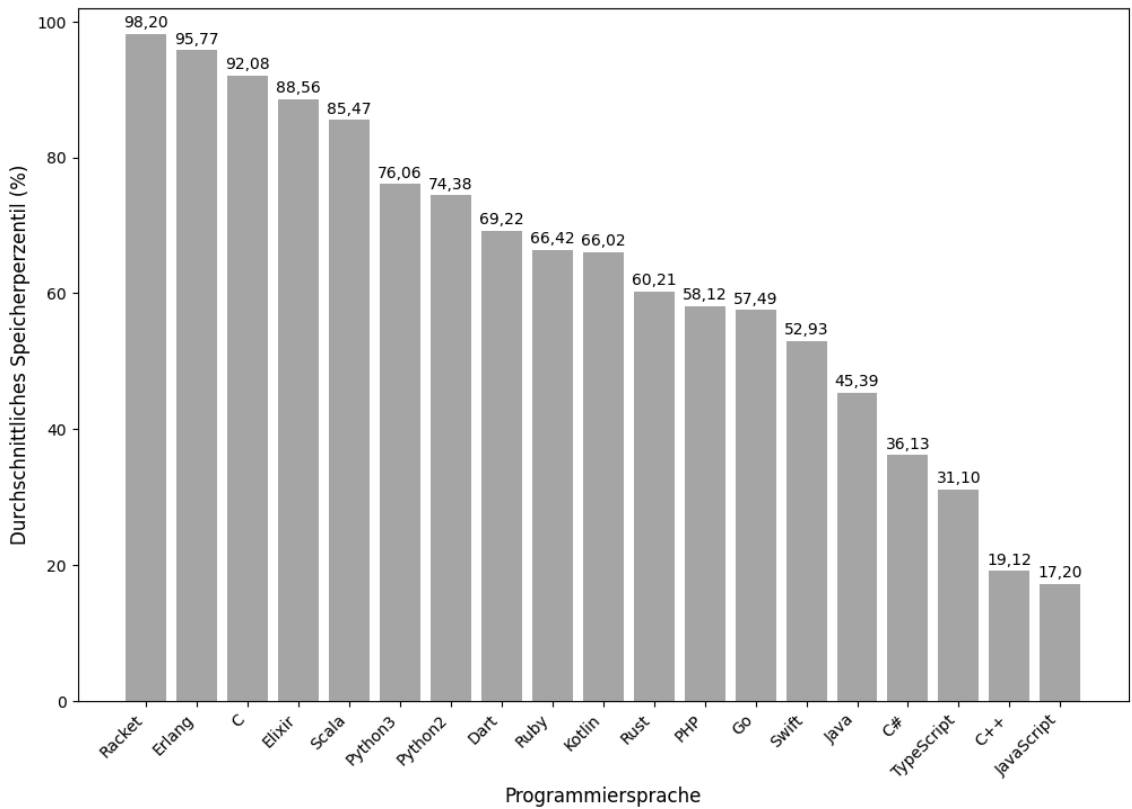

*Anmerkung.* Quelle: Eigene Darstellung

Die Ergebnisse zeigen eine signifikante Variabilität zwischen den einzelnen Programmiersprachen. In 14 der 19 untersuchten Sprachen erzielte ChatGPT 4 höhere Werte als 50 % der Nutzer auf LeetCode, während es in den restlichen fünf Sprachen unter diesem Median lag. Besonders hohe Perzentile wurden für die Programmiersprachen Racket, Erlang und Elixir beobachtet. Es kann jedoch angenommen werden, dass diese Ergebnisse, analog zu den bei der Laufzeiteffizienz erzielten Werten, verzerrt sind.

Im Gegensatz dazu erreichte ChatGPT 4 in C eine herausragende Leistung im 92. Perzentil. Ebenso erzielte ChatGPT 4 in der Programmiersprache Scala, ähnlich wie bei der Laufzeiteffizienz, eine starke Leistung im 89. Perzentil. Im Unterschied zu den Werten der Laufzeiteffizienz waren auch Python 3 und Python 2 mit Werten um das 75. Perzentil bemerkenswert, wobei die aktuellere Version in diesem Fall leicht vorne lag.

Im mittleren Bereich befanden sich die Programmiersprachen Dart, Ruby und Kotlin, gefolgt von Rust, PHP, Go und Swift. ChatGPT 4 wies in diesen Programmiersprachen im Vergleich zu anderen LeetCode-Nutzern weiterhin überdurchschnittliche Werte auf, die zwischen dem 70. und dem 52. Perzentil lagen. Ab diesem Punkt begannen die Leistungen zu fallen, wobei ChatGPT 4 für Aufgaben in Java durchschnittlich das 45. Perzentil, für C# das 36. Perzentil und für TypeScript das 31. Perzentil erreichte. In den Aufgaben zu C++ und JavaScript wurde lediglich das 19. beziehungsweise 17. Perzentil erreicht. Trotz der Tatsache, dass sowohl TypeScript als auch JavaScript in den unteren Rängen angesiedelt waren, wiesen sie dennoch signifikant unterschiedliche Werte auf. Dabei zeigte sich, dass ChatGPT 4 im Umgang mit JavaScript größere Schwierigkeiten im Speichermanagement hatte als bei TypeScript.

In Bezug auf das Abstraktionsniveau stellte sich heraus, dass ChatGPT 4 in Sprachen hoher Abstraktionsebene mit einem durchschnittlichen Perzentil von 64,07 besser abschnitt als in Sprachen niedriger Abstraktionsebene, wo im Durchschnitt ein Wert von 57,23 erzielt wurde. Nachdem die Programmiersprachen Racket, Erlang und Elixir, wie bereits bei der Bewertung der Laufzeiteffizienz, zugunsten eines präziseren Bildes außer Acht gelassen wurden, reduzierte sich der Wert für Programmiersprachen mit hoher Abstraktionsebene auf das 56,54. Perzentil. Im Gegensatz dazu blieb der Wert für Programmiersprachen mit niedriger Abstraktionsebene konstant im 57,23. Perzentil. Es zeigte sich somit eine minimale Tendenz zugunsten von Sprachen mit niedriger Abstraktionsebene.

Bei Betrachtung der Typisierung zeigten die Durchschnittswerte unter Einbeziehung von Racket, Erlang und Elixir eine signifikante Präferenz für dynamisch typisierte Sprachen, wo ein Perzentil von 71,84 erreicht wurde. Im Vergleich dazu erreichte ChatGPT 4 in statisch typisierten Sprachen das 55,92. Perzentil. Selbst unter Ausschluss der Programmiersprachen Racket, Erlang und Elixir lag der Wert für dynamisch typisierte Sprachen weiterhin im 58,44. Perzentil und war somit höher als der Wert für statisch typisierte Sprachen, der konstant im 55,92. Perzentil verblieb. ChatGPT 4 zeigte somit bezüglich der Speichereffizienz eine tendenzielle Präferenz für dynamisch typisierte Programmiersprachen.

# 5 Diskussion

Im folgenden Abschnitt werden die im Ergebnisteil präsentierten Resultate interpretiert und analysiert. Zunächst werden die Implikationen erörtert, die sich aus den Ergebnissen ableiten lassen. Danach werden die Limitationen der Studie aufgezeigt und abschließend Empfehlungen für zukünftige Forschungen gegeben.

## 5.1 Implikationen

Im Einklang mit den Erkenntnissen von Bucaioni et al. (2024) und Zhang et al. (2023) zeigte sich, dass ChatGPT 4 bei einfachen Aufgaben in allen Programmiersprachen durchschnittlich hohe Lösungsraten erzielte. Allerdings traten, im Gegensatz zu den Ergebnissen von Bucaioni et al. (2024), bereits ab dem mittleren Schwierigkeitsgrad signifikante Schwierigkeiten auf, sodass sich ChatGPT 4 bei komplexen Problemstellungen schließlich erheblichen Herausforderungen gegenübersah, wobei zwischen den verschiedenen Sprachen nur maximal zwei Probleme gelöst werden konnten. Dies verdeutlicht einen klaren Trend, wonach ChatGPT 4 in allen untersuchten Programmiersprachen bei steigendem Schwierigkeitsgrad auf zunehmend größere Probleme stößt und in keiner der Programmiersprachen in der Lage ist, effektiv Lösungen für komplexe Aufgaben bereitzustellen.

Bei der Betrachtung aller Programmiersprachen und Aufgaben ergab sich eine Gesamtlösungsrate von 39,67 %. Obwohl diese Rate als akzeptabel angesehen werden kann, liegt sie dennoch signifikant unter einem optimalen Kompetenzniveau. Bemerkenswert ist, dass ChatGPT 4 die meisten Aufgaben in der Programmiersprache Kotlin bewältigen konnte, wobei eine Lösungsrate von 45,21 % erzielt wurde. Im Gegensatz dazu wurden in den Programmiersprachen Elixir und Erlang mit einer Lösungsrate von lediglich 23,94 % die geringsten Werte erzielt. Es zeigt sich also, dass ChatGPT 4 zwischen verschiedenen Sprachen Leistungsschwankungen aufweist.

Die Erkenntnisse von Buscemi (2023) zeigen, dass das Vorgängermodell ChatGPT 3.5 in dynamisch typisierten Hochsprachen besser abschneidet als bei statisch typisierten Sprachen mit niedrigem Abstraktionsniveau. Diese Beobachtung konnte in der vorliegenden Bachelorarbeit für ChatGPT 4 nicht bestätigt werden. Im Gegenteil zeigt sich, dass ChatGPT 4 eine Präferenz für Sprachen mit niedrigem Abstraktionsniveau bei der Lösung von Aufgaben aufweist und statisch typisierte Sprachen bevorzugt.

Bei weiterer Analyse der Gesamtlösungsraten war für die ersten 16 Programmiersprachen ein Cluster mit ähnlichen Werten erkennbar, wobei es trotz der engen Werteverteilung dennoch geringfügige Unterschiede zwischen den einzelnen Sprachen gab. Innerhalb dieses Clusters wurde in Kotlin die höchste Anzahl gelöster Probleme (85) und in C die geringste Anzahl (75) registriert.

Diese Varianz könnte einerseits auf die unterschiedliche Verfügbarkeit einer größeren Menge sowie qualitativ hochwertigerer Trainingsdaten zurückzuführen sein. Andererseits ist es ebenso möglich, dass durch das nicht-deterministische Verhalten von ChatGPT 4, bei dem dieselbe Frage unterschiedlich beantwortet werden kann, zufällig in einigen Sprachen mehr Probleme gelöst werden konnten als in anderen (OpenAI, o. D.-a). Es ist somit denkbar, dass ChatGPT 4 in bestimmten Sprachen bei der ersten Iteration zufällig auf komplexere Fehler stieß, die in zwei weiteren Versuchen nicht behoben werden konnten, jedoch in späteren Iterationen potenziell hätten korrigiert werden können. Zur Überprüfung dieser Hypothese sollten zukünftige Studien eine größere Anzahl von Iterationen und Programmierproblemen berücksichtigen, um zu bestimmen, ob die Werte konvergieren, konstant bleiben oder divergieren.

Eine weiterführende Betrachtung der Gesamtlösungsraten offenbarte ein zweites Cluster, das sich durch signifikant niedrigere Lösungsraten im Vergleich zu den Programmiersprachen des ersten Clusters auszeichnete und die weniger verbreiteten Programmiersprachen Racket, Erlang und Elixir umfasste. Wie Buscemi (2023) in seinen Untersuchungen bereits zum Vorgängermodell ChatGPT 3.5 feststellte, scheint die Popularität einer Programmiersprache einen signifikanten Einfluss auf die Code-Generierungsfähigkeit von ChatGPT 4 auszuüben. Dies ist vermutlich auf die größere Verfügbarkeit von Trainingsdaten für weiter verbreitete Programmiersprachen im Vergleich zu weniger bekannten Sprachen zurückzuführen, was die Entstehung der zwei Cluster, bestehend aus populären und weniger populären Sprachen, erklären würde (*Stack Overflow Developer Survey 2023*, 2023; Cass, 2023; O'Grady, 2024).

Bei der Analyse der aufgetretenen Fehler ließen sich diese beiden Cluster ebenfalls beobachten. Im ersten Cluster war der Fehler *Wrong Answer* am häufigsten zu erkennen, der auftrat, wenn nicht jeder der mehreren hundert Testfälle auf LeetCode bestanden wurde. Dies zeigt, dass ChatGPT 4 dazu in der Lage ist, in diesen Programmiersprachen in den meisten Fällen syntaktisch einwandfreie und zur Laufzeit fehlerfreie Funktionen zu erstellen, jedoch hinsichtlich ihrer Korrektheit scheitert.

Es zeigte sich jedoch ebenso, dass ChatGPT 4 in vielen Instanzen der untersuchten Sprachen trotz des *Wrong Answer*-Fehlers eine erhebliche Anzahl von Testfällen erfolgreich passieren konnte. Daraus lässt sich schließen, dass die Lösung möglicherweise akzeptiert worden wäre, wenn weniger Testfälle vorgelegen hätten, was jedoch zu einer suboptimalen Funktion geführt hätte. Dies unterstreicht erneut die Bedeutung einer gründlichen Überprüfung des von ChatGPT 4 generierten Codes mittels umfangreicher Testfälle. Dies könnte in einigen Fällen jedoch potenziell schwierig sein, insbesondere wenn es nicht möglich ist, den von ChatGPT 4 entwickelten Algorithmus selbst vollständig nachzuvollziehen. Die von Zhang et al. (2023) identifizierte Unfähigkeit zur Selbstvalidierung von ChatGPT 4 stellt hierbei den kritischen Punkt dar.

Im zweiten Cluster, das weniger populäre Programmiersprachen umfasste, waren hingegen auffällig viele Compiler- und Laufzeitfehler festzustellen. Insbesondere traten hier grundlegende syntaktische Missverständnisse in hoher Anzahl auf. Beispiele für diese Problematik waren das Vergessen, Klammern zu schließen, sowie das versehentliche Schließen von Klammern mit eckigen statt runden Klammern. Darüber hinaus wurden Versuche beobachtet, bestimmte Operationen durchzuführen, die in populären

Programmiersprachen üblich sind, jedoch in Elixir, Erlang und Racket keine Anwendung finden.

Dies deutet darauf hin, dass ChatGPT 4 in einigen Fällen Schwierigkeiten hat, die spezifische Syntax dieser Sprachen einzuhalten und sie präzise von anderen Sprachen zu unterscheiden. Dabei nimmt das Modell an, dass Operationen, die in anderen Sprachen verwendet werden, auch in diesen Sprachen anwendbar sind. Ein solches Verhalten weist auf signifikante Lücken im Verständnis der strukturellen Eigenschaften dieser Sprachen hin, was höchstwahrscheinlich ebenfalls auf die begrenzte Verfügbarkeit von Trainingsdaten aufgrund ihrer geringeren Popularität zurückzuführen ist.

Darüber hinaus war festzustellen, dass auch in den Sprachen des ersten Clusters, wenngleich in geringerem Umfang, signifikante Laufzeit- und Compilerfehler auftraten, wobei insbesondere der hohe Anteil an Laufzeitfehlern bei der Programmiersprache Dart hervorstach. Diese Fehler resultierten ebenfalls aus einem grundlegenden Missverständnis der Merkmale der jeweiligen Sprachen. Dies erfordert notwendigerweise das Eingreifen menschlicher Programmierer, um den Code angemessen zu korrigieren oder ChatGPT 4 auf die begangenen Fehler hinzuweisen, und verdeutlicht, dass ChatGPT 4 auch in diesen Sprachen derzeit nicht makellos arbeitet und der generierte Code daher nicht in jedem Fall uneingeschränkt übernommen werden sollte. ChatGPT 4 sollte daher aktuell in allen Sprachen als unterstützendes Werkzeug betrachtet werden, wobei auch signifikante Fehler auftreten können.

Im Hinblick auf die anderen aufgetretenen Fehler zeigte sich, dass ChatGPT 4 auf den Fehler *Time Limit Exceeded* in den meisten Programmiersprachen proportional gleich häufig stieß, und zwar stets bei denselben Aufgaben mittlerer und hoher Schwierigkeit. Dies deutet darauf hin, dass ChatGPT 4 diesen Fehler weitgehend unabhängig von der verwendeten Programmiersprache begeht.

Eine signifikante Ausnahme in diesem Zusammenhang stellte die Programmiersprache Scala dar. Während in Scala weniger Fehler dieser Art auftraten, wurde im Vergleich dazu jedoch der ansonsten relativ seltene Fehler *Memory Limit Exceeded* verhältnismäßig häufig beobachtet. Diese Beobachtung erscheint paradox, da ChatGPT 4 in Scala hohe Werte der Speichereffizienz erreichte. Dies lässt darauf schließen, dass es eine große Variabilität in den Ergebnissen gibt. Entweder gelingt es ChatGPT 4, die speziellen Eigenschaften Scalas zu nutzen, um speichereffiziente Lösungen zu entwickeln, oder es scheitert grundsätzlich an diesem Vorhaben.

In Bezug auf die Qualität des Codes wurden die Metriken der Laufzeit- und Speichereffizienz verwendet. Es ist wichtig zu betonen, dass diese beiden Metriken nicht ausreichen, um die Qualität des Codes vollständig zu erfassen. Weitere wichtige Aspekte wie Wartbarkeit, Lesbarkeit und Erweiterbarkeit, die ebenfalls zur Codequalität beitragen, wurden in dieser Bachelorarbeit nicht behandelt, könnten jedoch in weiterführender Forschung von Bedeutung sein.

Bezüglich der Laufzeiteffizienz demonstrierte ChatGPT 4 die Fähigkeit, in allen untersuchten Programmiersprachen niedrigere Laufzeitwerte als der durchschnittliche LeetCode-Nutzer zu erzielen. Dies deutet auf eine ausgeprägte Kompetenz des Modells hin, Funktionen mit vergleichsweise minimaler Laufzeit in den betrachteten Programmiersprachen zu implementieren.

Die Streuung zwischen den Programmiersprachen war bei dieser Metrik erheblich, was darauf hinweist, dass, obwohl ChatGPT 4 in jeder Sprache bessere Ergebnisse erzielte als der durchschnittliche LeetCode-Nutzer, in einigen Sprachen hinsichtlich der Laufzeiteffizienz nochmals signifikant höhere Kompetenzen zu erwarten sind als in anderen. Insbesondere bei den Programmiersprachen Racket, Erlang und Elixir, in denen ChatGPT 4 insgesamt eine geringe Lösungsrate aufwies, wurden hohe Werte festgestellt. Diese hohen Werte könnten darauf hindeuten, dass das Modell über eine herausragende Fähigkeit verfügt, in diesen Sprachen laufzeiteffizienten Code zu erzeugen. Allerdings ist zu berücksichtigen, dass in diesen Sprachen generell relativ wenige Probleme gelöst wurden, die Anzahl an Lösungen anderer Nutzer in diesen Sprachen durchschnittlich gering war und viele der verwendeten Probleme relativ neu waren. Aufgrund der daraus resultierenden reduzierten Vergleichsdatenmenge sind die Werte dieser Sprachen höchstwahrscheinlich nicht repräsentativ.

Weiterhin zeigten sich im oberen Quartil jedoch hohe Werte für die Sprachen Scala und PHP. Auch in Java und C wurden Werte erzielt, die nur geringfügig hinter den besten 25 % aller Lösungen zurückblieben. Diese Ergebnisse deuten auf die Fähigkeit von ChatGPT 4 hin, in diesen Programmiersprachen relativ laufzeiteffizienten Code zu generieren. Die Werte fielen bis auf das 52. Perzentil für die Programmiersprache JavaScript, was jedoch weiterhin über dem Durchschnitt aller LeetCode-Nutzer lag und somit als solide Leistung betrachtet werden kann.

Insgesamt zeigen die Ergebnisse zur Laufzeiteffizienz, dass ChatGPT 4 dazu neigt, in statisch typisierten Sprachen sowie in Sprachen mit niedrigem Abstraktionsniveau laufzeiteffizienteren Code zu generieren. Dies impliziert, dass ChatGPT 4 die strukturellen und typensicheren Eigenschaften dieser Sprachen besser nutzen kann und dadurch Optimierungen ermöglicht, die in dynamisch typisierten oder höher abstrahierten Sprachen schwieriger umzusetzen sind.

Es ist zudem bemerkenswert, dass ChatGPT 4, ähnlich wie bei der Lösungsrate in Python 2, auch hinsichtlich der Laufzeiteffizienz unerwartet bessere Leistungen erzielte als in Python 3. Die Ursachen für diese Diskrepanz sind nicht unmittelbar ersichtlich. Es ist möglich, dass für Python 2 eine größere Menge an Trainingsdaten verfügbar war, was die Entwicklung laufzeiteffizienter und fehlerfreier Lösungen begünstigte. In weiterführenden wissenschaftlichen Untersuchungen sollte diese Erkenntnis durch die Einbeziehung einer größeren Anzahl von Programmierproblemen überprüft werden, um festzustellen, ob es sich nicht um ein zufälliges Ergebnis handelt, das durch das nicht-deterministische Verhalten von ChatGPT 4 bedingt ist. Zusätzlich wäre es ratsam, diese Tendenz auch zwischen verschiedenen Versionen anderer Programmiersprachen zu überprüfen.

Bei der Betrachtung der zweiten Metrik zur Bewertung der Codequalität, der Speichereffizienz, offenbarte sich eine noch größere Varianz zwischen den Werten der verschiedenen Programmiersprachen im Vergleich zur Laufzeiteffizienz. Dies impliziert, dass die Wahl der Programmiersprache einen signifikanten Einfluss auf die Speichereffizienz des von ChatGPT 4 generierten Codes hat. Bei der Auswahl einer Programmiersprache für Projekte, bei denen effizientes Speichermanagement erforderlich ist und ChatGPT 4 zum Einsatz kommt, sollte dieser Aspekt berücksichtigt werden.

Es wurde festgestellt, dass ChatGPT 4 in 14 der 19 untersuchten Programmiersprachen Werte erzielte, die über dem Median lagen, was auf eine relative Stärke des Modells in diesen Sprachen hinweist. Insbesondere zeichneten sich erneut die Sprachen Racket, Erlang und Elixir aus, wobei die bereits bei der Laufzeiteffizienz erwähnte Verzerrung auch hier eine Rolle spielte. Des Weiteren waren auf den oberen Plätzen auffällig hohe Werte für C sowie für die bereits bei der Laufzeiteffizienz sehr starke Programmiersprache Scala zu beobachten, was darauf hinweist, dass ChatGPT 4 in diesen Sprachen besonders überdurchschnittliche Kompetenzen zur Generierung speichereffizienten Codes aufweist. Auf der anderen Seite jedoch zeigte sich, dass in fünf Sprachen die Werte unter dem Median lagen, wobei vor allem die besonders auffällig schwachen Werte für C++ und JavaScript herausstachen. Diese Werte deuten darauf hin, dass ChatGPT 4 in diesen Sprachen erhebliche Schwierigkeiten beim Speichermanagement aufweist.

Zum einen ist festzustellen, dass ChatGPT 4 eine Präferenz für die Generierung speichereffizienten Codes in Programmiersprachen mit niedrigem Abstraktionsgrad aufweist, wobei diese Tendenz jedoch nur marginal ausgeprägt ist. Zum anderen wurden auffällige Tendenzen bezüglich der Typisierung beobachtet. Es wurde festgestellt, dass ChatGPT 4 auch im JavaScript-Superset TypeScript, ähnlich wie bereits in JavaScript, unterdurchschnittliche Leistungen im Vergleich zu anderen LeetCode-Nutzern zeigte, jedoch in TypeScript dennoch signifikant höhere Werte als in JavaScript erreichte. Eine plausible Erklärung hierfür könnte sein, dass ChatGPT 4 durch die statische Typisierung von TypeScript eine effizientere Speicherverwaltung ermöglicht wird.

Das Gesamtbild betrachtend, ist jedoch ersichtlich, dass ChatGPT 4 eine unerwartete Präferenz für dynamisch typisierte Sprachen hinsichtlich der Speichereffizienz aufweist. Es wäre daher naheliegend anzunehmen, dass im Fall von TypeScript und JavaScript die statische Typisierung tatsächlich den Unterschied ausmacht, wobei im Allgemeinen eine Präferenz für dynamische Typisierung zu bestehen scheint.

Diese Tendenz könnte damit erklärt werden, dass ChatGPT 4 die Speicheroptimierungen, die durch statische Typisierung ermöglicht werden, aufgrund fehlenden Verständnisses nicht vollständig nutzen kann, was zu relativ speicherineffizienten Lösungen führt. Dynamisch typisierte Sprachen erfordern auf der anderen Seite in der Regel weniger spezialisierte Kenntnisse für die Speicheroptimierung, da viele dieser Aspekte oftmals von der Laufzeitumgebung übernommen werden. Die spezifischen Gründe für diese unvorhergesehene Präferenz sind jedoch nicht unmittelbar ersichtlich und bedürfen einer detaillierteren Untersuchung in zukünftigen Forschungen.

## 5.2 Limitationen

Trotz der umfassenden Sorgfalt, die bei der Datenerhebung und -auswertung für diese Arbeit angewandt wurde, sind im Folgenden einige potenzielle Limitationen zu erörtern, die die Validität und Generalisierbarkeit der Ergebnisse beeinflussen könnten.

Zunächst ist festzuhalten, dass trotz der Angabe von OpenAI, ChatGPT 4 verfüge über einen Wissensstand bis April 2023, nicht vollständig ausgeschlossen werden kann, dass das Modell auch Informationen über diesen Zeitpunkt hinaus besitzt (OpenAI, 2023b). Diese Annahme lässt sich dadurch stützen, dass ChatGPT 4 nach Angaben von OpenAI kontinuierlich durch Nutzerinteraktionen trainiert wird, was potenziell dazu führen könnte, dass das Modell über Aufgaben und Lösungen informiert ist, die über den angegebenen Wissensstand hinausgehen (OpenAI, o. D.-b).

Darüber hinaus ist es möglich, dass selbst ein relativ umfangreicher Datensatz von 188 Problemen noch immer zu klein ist, um allgemeingültige Aussagen zu treffen. Zudem könnte eine andere oder breitere Auswahl an Programmiersprachen zu unterschiedlichen Ergebnissen führen. Weiterhin ist vorstellbar, dass die Lösungsrate potenziell höher ausgefallen wäre, wenn mehr als drei Iterationen zur Problemlösung durch Feedback ermöglicht worden wären. Dies hätte jedoch eine erhebliche Zunahme des Zeitaufwands zur Folge gehabt, die den Rahmen dieser Bachelorarbeit deutlich überschritten hätte.

Des Weiteren ist zu berücksichtigen, dass für einige der neueren Probleme nur begrenzt Vergleichsdaten zur Verfügung standen, da nur wenige Lösungen anderer Nutzer vorlagen, insbesondere in Bezug auf weniger verbreitete Programmiersprachen. Dies führte dazu, dass die entsprechenden Werte zur Laufzeit- und Speichereffizienz dieser Sprachen ein verzerrtes Bild abgaben. Zudem ist es in diesem Zusammenhang weiterhin wichtig zu betonen, dass der durchgeführte Vergleich nicht als Vergleich mit menschlichen Nutzern missverstanden werden sollte. Es kann nicht davon ausgegangen werden, dass die Vergleichsdaten ausschließlich von menschlichen Programmierern stammen. Dies wird unter anderem dadurch verdeutlicht, dass ChatGPT 4 und andere LLMs bereits, wie in Abschnitt 2.2 erwähnt, in einigen Studien involviert sind, die, ähnlich wie diese Bachelorarbeit, LeetCode als Plattform zur Bewertung von Code-Lösungen nutzen.

Zudem ist klarzustellen, dass die in dieser Arbeit verwendeten Aufgaben auf die Implementierung einer einzelnen Funktion beschränkt sind. Dies verdeutlicht zwar die Fähigkeit von ChatGPT 4, Code für isolierte Funktionen zu generieren, sollte jedoch nicht fälschlicherweise als umfassende Fähigkeit zur gesamten Softwareentwicklung interpretiert werden. Letztere erfordert die Integration verschiedener Komponenten unter Anwendung spezifischer Entwurfsmuster und ist somit von weitaus höherer Komplexität. Die Untersuchung konzentriert sich ausschließlich auf die Implementierung spezifischer Funktionen und umfasst nicht den gesamten Softwareentwicklungsprozess, der zusätzliche Schritte wie Planung, Entwurf, Integration, Testen, Aktualisierung und Erweiterung der Software beinhaltet. Aus diesem Grund wird auch strikt von Code-Generierung und nicht von Programmierung gesprochen.

Weiterhin lässt sich feststellen, dass die Prozesse zur Ermittlung der Laufzeit- und Speicherwerte, die zur Analyse der Codequalität mit den Werten anderer Nutzer verglichen wurden, sowie der Kategorisierung der Aufgaben nach Schwierigkeitsgrad von LeetCode nicht offengelegt wurden. Daher bleibt unklar, ob die bereitgestellten Werte und Einteilungen tatsächlich präzise und zuverlässig sind.

Abschließend sei darauf hingewiesen, dass zur Bewertung der Codequalität die Metriken der Laufzeit- und Speichereffizienz allein nicht ausreichen. Zusätzlich können Faktoren wie Wartbarkeit, Skalierbarkeit, Lesbarkeit, Fehlerbehandlung, Dokumentation und die nach McCabe definierte zyklomatische Komplexität zur Bewertung der Codequalität herangezogen werden.

## 5.3 Empfehlung für zukünftige Forschung

Zunächst sollte betont werden, dass zukünftige Forschungen erheblich von der Nutzung einer größeren Anzahl von Programmieraufgaben zur Datenerhebung und der anschließenden Analyse profitieren würden. Es wäre zudem empfehlenswert, die Anzahl der Iterationen über drei hinaus zu erhöhen, um sicherzustellen, dass ChatGPT 4 eine ausreichende Anzahl von Versuchen erhält, sich selbst zu korrigieren. Durch die daraus resultierende präzisere Bestimmung der Lösungsraten könnte festgestellt werden, ob die Werte konvergieren, konstant bleiben oder divergieren. Spezifisch könnte dahingehend zusätzlich überprüft werden, ob ChatGPT 4 tatsächlich eine höhere Kompetenz in der älteren Version 2 im Vergleich zur aktuellen Version 3 von Python aufweist. Dabei wäre es ratsam, diese Tendenz auch im Kontext verschiedener Versionen anderer Programmiersprachen zu analysieren. Bei der Auswahl der zu untersuchenden Probleme sollte darauf geachtet werden, dass diese nicht allzu neu sind. Dies stellt sicher, dass genügend Lösungen von anderen Nutzern vorliegen, wodurch die Validität eines Vergleichs der Laufzeit- und Speicherwerte gewährleistet wird.

Des Weiteren könnten zusätzliche Programmiersprachen in die Analyse einbezogen werden, insbesondere solche, die weniger verbreitet sind. Dies würde es ermöglichen, Unterschiede zwischen weniger populären Sprachen zu untersuchen und festzustellen, ob sich die Fähigkeiten von ChatGPT 4 zwischen diesen nochmals grundlegend unterscheiden.

Weiterhin könnte die Analyse durch mehrere Maßnahmen erweitert werden. Zum einen wäre es vorteilhaft, die spezifischen Themen der Programmierprobleme zu berücksichtigen und zu untersuchen, ob es Leistungsunterschiede seitens ChatGPT 4 zwischen den Themenbereichen und Programmiersprachen gibt. Darüber hinaus sollte analysiert werden, wie viele Versuche ChatGPT 4 benötigt, um das Problem in verschiedenen Sprachen erfolgreich zu lösen. Zielführend wäre zudem, die Kompetenz von ChatGPT 4 bei der Interpretation und Zusammenfassung sowie beim Debuggen von Code in unterschiedlichen Programmiersprachen zu überprüfen, um zusätzliche Leistungsunterschiede festzustellen. Dahingehend wäre es zusätzlich sinnvoll, ChatGPT 4 in verschiedenen Programmiersprachen komplexeren und praxisbezogenen Code entwerfen und implementieren zu lassen. Dabei sollte geprüft werden, ob das Modell gängige Designmuster und Best Practices der Sprachen korrekt und effizient anwendet.

Abschließend wird empfohlen, auch zukünftige Iterationen von ChatGPT 4 sowie andere LLMs in die Untersuchung einzubeziehen, um deren Fähigkeiten in der Code-Generierung zu überprüfen und den aktuellen Stand der LLMs in verschiedenen Programmiersprachen zu evaluieren.

# 6 Fazit

In dieser Bachelorarbeit wurden die Code-Generierungsfähigkeiten von ChatGPT 4 in 19 Programmiersprachen untersucht. Die ausgewählten Sprachen umfassten Kotlin, Java, Rust, Scala, Go, C#, C++, PHP, Python (Versionen 2 und 3), TypeScript, Dart, JavaScript, Swift, Ruby, C sowie Racket, Elixir und Erlang. Dabei wurden 188 Programmierprobleme unterschiedlicher Schwierigkeitsgrade von der Coding-Interview-Plattform LeetCode verwendet. Für jedes Problem und jede Sprache wurden spezifische Prompts an ChatGPT 4 übermittelt und die generierten Codes auf LeetCode hinsichtlich Funktionalität und Effizienz evaluiert. ChatGPT 4 hatte dabei drei Versuche, korrekten Code bereitzustellen, wobei nach jedem fehlerhaften Versuch die entsprechenden Fehlermeldungen zur Verfügung gestellt wurden, damit das Modell den Code korrigieren konnte. Die Datenerhebung und -analyse wurden durch ein eigens entwickeltes Python-Projekt automatisiert.

Die Ergebnisse zur ersten Forschungsfrage (F1) zeigten, dass ChatGPT 4 insgesamt 39,67 % der Probleme über alle untersuchten Programmiersprachen und Aufgaben erfolgreich lösen konnte, was auf ein akzeptables, jedoch noch nicht optimales Kompetenzniveau hinweist. Bei einfachen Aufgaben erreichte ChatGPT 4 eine hohe Erfolgsquote von 85,07 %, während sich bei mittelschweren Aufgaben eine signifikant reduzierte Lösungsrate von 28,56 % zeigte und bei schwierigen Aufgaben nur eine marginale Erfolgsrate von 2,43 % erzielt wurde. Diese Resultate verdeutlichen, dass ChatGPT 4 aktuell Schwierigkeiten hat, Aufgaben zunehmender Komplexität in den untersuchten Programmiersprachen erfolgreich zu bewältigen. Es zeigt sich zudem, dass ChatGPT 4 bei der Lösung von Problemen eine Präferenz für Sprachen mit niedrigem Abstraktionsniveau sowie für Sprachen mit statischer Typisierung aufweist.

Eine Analyse der einzelnen Gesamtlösungsraten ergab, dass ChatGPT 4 in verschiedenen Programmiersprachen unterschiedlich erfolgreich war. In Kotlin wurde mit 85 gelösten Aufgaben das beste Ergebnis erzielt, während in Erlang und Elixir nur 45 Aufgaben gelöst wurden, was den geringsten Wert darstellt. Es wurden zwei Cluster identifiziert. Der erste Cluster umfasste die 16 verbreitetsten Sprachen, wobei ChatGPT 4 in Kotlin mit 85 gelösten Aufgaben das beste und in C mit 75 gelösten Aufgaben das schlechteste Ergebnis dieser Gruppe erzielte. Der zweite Cluster beinhaltete die weniger verbreiteten Programmiersprachen Elixir, Erlang und Racket mit signifikant niedrigeren Lösungsraten im Vergleich zum ersten Cluster. Dies verdeutlicht, dass die Fähigkeit von ChatGPT 4 zur Lösung von Programmierproblemen je nach Programmiersprache variiert. In weniger populären Programmiersprachen zeigt ChatGPT 4 geringere Kompetenzen, was vermutlich auf eine geringere Anzahl und niedrigere Qualität der Trainingsdatensätze zurückzuführen ist.

Trotz der engen Werteverteilung der Programmiersprachen im ersten Cluster bestehen dennoch geringfügige Unterschiede zwischen ihnen. Diese Unterschiede können einerseits auf die Anzahl und Qualität der Trainingsdatensätze sowie andererseits auf das nicht-deterministische Verhalten von ChatGPT 4 zurückgeführt werden. Im Hinblick auf letztere Möglichkeit könnten in der ersten Iteration zufällig komplexere Fehler aufgetreten sein, die in zwei weiteren Versuchen nicht behoben werden konnten, aber in weiteren Iterationen möglicherweise gelöst worden wären. Daher wird empfohlen, die Anzahl der Iterationen in zukünftiger Forschung zu erhöhen und mehr Probleme in die Analyse einzubeziehen, um die Konsistenz der Werte zu überprüfen.

Bei der Analyse der aufgetretenen Fehler zur Beantwortung der zweiten Forschungsfrage (F2) ließen sich die beiden Cluster ebenfalls deutlich erkennen. Im ersten Cluster überwog der Fehler *Wrong Answer*, was darauf hinweist, dass Fehler im Code von ChatGPT 4 in diesen Sprachen meistens darauf zurückzuführen sind, dass die spezifischen Anforderungen nicht vollständig erfüllt werden, der Code jedoch syntaktisch korrekt und zur Laufzeit fehlerfrei ist.

Im zweiten Cluster der Sprachen Elixir, Erlang und Racket traten überwiegend Compiler- und Laufzeitfehler auf, die hauptsächlich auf syntaktische Missverständnisse und Verwechslungen mit Eigenschaften anderer Sprachen zurückzuführen waren. Dies weist darauf hin, dass ChatGPT 4 teilweise Schwierigkeiten hat, die Syntax dieser Sprachen korrekt einzuhalten und sie von anderen zu unterscheiden, was möglicherweise erneut auf begrenzte Trainingsdaten für diese Sprachen zurückzuführen ist.

Im Cluster der weiter verbreiteten Sprachen wurden ebenfalls Compiler- und Laufzeitfehler festgestellt, insbesondere eine hohe Anzahl an Laufzeitfehlern bei Dart. Dies zeigt, dass ChatGPT 4 auch in diesen Sprachen nicht immer fehlerfreie Lösungen liefert und der generierte Code daher nicht ohne weitere Prüfung übernommen werden sollte.

Der *Time Limit Exceeded*-Fehler trat bei Aufgaben mittleren und schweren Schwierigkeitsgrades proportional gleich häufig auf. Dies deutet darauf hin, dass ChatGPT 4 diesen Fehler weitgehend unabhängig von der verwendeten Programmiersprache begeht. Eine Ausnahme bildet Scala, wo viele *Memory Limit Exceeded*-Fehler auftraten, die zwischen den Sprachen generell selten waren. Dies ist bemerkenswert, da ChatGPT 4 in Scala üblicherweise speichereffiziente Lösungen lieferte, was darauf hinweist, dass das Modell entweder speichereffiziente Lösungen in Scala bereitstellt oder grundsätzlich daran scheitert.

In Bezug auf die dritte Forschungsfrage (F3) zeigte sich, dass ChatGPT 4 hinsichtlich der Laufzeiteffizienz in allen betrachteten Programmiersprachen bessere Lösungen als der durchschnittliche LeetCode-Nutzer erzielt hat, wobei die Streuung der Ergebnisse zwischen den verschiedenen Sprachen erheblich war. Die Verteilung erstreckte sich von Racket, Erlang und Elixir, bei denen sehr hohe Ergebnisse im Bereich des 98. bis 88. Perzentils erzielt wurden, die jedoch vermutlich einer Verzerrung unterlagen, bis hin zu Python 3 und JavaScript, bei denen ChatGPT 4 Ergebnisse erzielte, die geringfügig über dem 50. Perzentil lagen. Diese Ergebnisse zeigen, dass ChatGPT 4 im Allgemeinen fähig ist, in allen untersuchten Programmiersprachen überdurchschnittlich laufzeiteffizienten Code zu generieren, wobei dennoch signifikante Unterschiede in der Kompetenz zwischen den verschiedenen Sprachen bestehen. Zudem zeigt ChatGPT 4 eine Tendenz, in statisch typisierten Sprachen sowie in Sprachen mit niedrigem Abstraktionsniveau laufzeiteffizienteren Code zu generieren als in dynamisch typisierten Sprachen und Sprachen mit hohem Abstraktionsniveau. Dies könnte auf eine effektivere Nutzung der strukturellen und typensicheren Eigenschaften dieser Sprachen sowie auf die dadurch möglichen Optimierungen zurückzuführen sein.

Hinsichtlich der Speichereffizienz zeigte sich eine noch größere Streuung. In 14 von 19 untersuchten Programmiersprachen lag ChatGPT 4 über dem Durchschnitt, in den übrigen fünf darunter. Auffällig hohe Werte wurden erneut für Racket, Erlang und Elixir festgestellt, wobei diese Werte wahrscheinlich erneut einer Verzerrung unterlagen. Signifikant erhöhte Perzentile wurden ebenfalls für die Programmiersprachen C und Scala festgestellt, was darauf hindeutet, dass von ChatGPT 4 in diesen Sprachen relativ speichereffizienter Code zu erwarten ist. Im Gegensatz dazu erzielte ChatGPT 4 in den Sprachen Java, C#, TypeScript, C++ und JavaScript Werte unter dem Median. Insbesondere in C++ und JavaScript erreichte das Modell lediglich das 19. bzw. 17. Perzentil. Dies zeigt, dass das Modell in diesen Sprachen Schwierigkeiten bei der Generierung speichereffizienten Codes aufweist.

ChatGPT 4 weist in Bezug auf die Speichereffizienz eine Präferenz für Programmiersprachen mit niedrigem Abstraktionsniveau sowie unerwartet für dynamisch typisierte Sprachen auf. Dies könnte darauf zurückzuführen sein, dass das Modell möglicherweise die Speicheroptimierungspotenziale statisch typisierter Programmiersprachen aufgrund eines unzureichenden Verständnisses nicht vollständig ausschöpfen kann, was zu vergleichsweise ineffizienten Lösungen führt. Dynamisch typisierte Programmiersprachen erfordern hingegen weniger spezialisierte Kenntnisse der Speicheroptimierung, da viele dieser Aspekte von der Laufzeitumgebung verwaltet werden.

Für zukünftige Forschung empfiehlt es sich, die Anzahl der Programmieraufgaben und Iterationen zu erhöhen, um präzisere Lösungsraten zu erzielen. Darüber hinaus könnten zusätzliche, weniger verbreitete Sprachen in die Untersuchung einbezogen werden, um mögliche Leistungsunterschiede zwischen diesen aufzuzeigen. Es wird zudem empfohlen, die spezifischen Themen der Programmieraufgaben sowie zusätzliche Qualitätsmetriken, die über die Laufzeit- und Speichereffizienz hinausgehen, zu berücksichtigen. Ferner sollte die Kompetenz von ChatGPT 4 in den Bereichen Interpretation, Zusammenfassung und Debugging von Code sowie beim Entwurf praxisbezogener, komplexer Codes überprüft werden. Abschließend sollten auch zukünftige Versionen von ChatGPT und anderen LLMs untersucht werden, um deren Fähigkeiten in der Code-Generierung weiter zu evaluieren.

# 7 Literaturverzeichnis

Anthropic. (2024, 4. März). Introducing the next generation of Claude. *anthropic.com*.

    Abgerufen am 12. Juni 2024, von https://www.anthropic.com/news/claude-3-family

Apple Inc. (o. D.). *Swift - Apple Developer*. Abgerufen am 19. März 2024, von

    https://developer.apple.com/swift/

Armstrong, J. (2007). A history of Erlang. In *Proceedings of the third ACM SIGPLAN*

    *conference on History of programming languages* (S. 6-1-6–26).

    https://doi.org/10.1145/1238844.1238850

Arnold, K., Gosling, J. & Holmes, D. (2005). *Java(TM) Programming Language, the (4th*

    *Edition)*. Addison-Wesley Professional. https://dl.acm.org/citation.cfm?id=1051069

Balbaert, I. (2012). *The Way to Go: A Thorough Introduction to the Go Programming*

    *Language*. IUniverse.

    https://cdn.conceptians.org/Computer%20Science(2)/Go%20programming/The%20W

    ay%20To%20Go_%20A%20Thorough%20Introduction%20To%20The%20Go%20Pr

    ogramming%20Language%20(%20PDFDrive%20).pdf

Belchin, M. & Juberias, P. (2015). Web Programming with Dart. In *Apress eBooks*. Apress.

    https://doi.org/10.1007/978-1-4842-0556-3

Bubeck, S., Chandrasekaran, V., Eldan, R., Gehrke, J., Horvitz, E., Kamar, E., Lee, P., Lee, Y.

    T., Li, Y., Lundberg, S., Nori, H., Palangi, H., Ribeiro, M. T. & Zhang, Y. (2023).

    Sparks of Artificial General Intelligence: Early experiments with GPT-4. *arXiv*

    *Preprint arXiv:2303.12712*, 21–26. https://doi.org/10.48550/arxiv.2303.12712

Bucaioni, A., Ekedahl, H., Helander, V. & Nguyen, P. T. (2024). Programming with ChatGPT: How far can we go? *Machine Learning With Applications*, *15*, 1–10. https://doi.org/10.1016/j.mlwa.2024.100526

Buscemi, A. (2023). A Comparative Study of Code Generation using ChatGPT 3.5 across 10 Programming Languages. *arXiv Preprint arXiv:2308.04477*, 1–12. https://doi.org/10.48550/arxiv.2308.04477

Cass, S. (2023, 29. August). *The Top Programming Languages 2023*. IEEE Spectrum. Abgerufen am 6. Juni 2024, von https://spectrum.ieee.org/the-top-programming-languages-2023

*ChatGPT*. (o. D.). Abgerufen am 19. März 2024, von https://chat.openai.com/?model=gpt-4

Chen, M., Tworek, J., Jun, H., Yuan, Q., De Oliveira Pinto, H. P., Kaplan, J., Edwards, H., Burda, Y., Joseph, N., Brockman, G., Ray, A., Puri, R., Krueger, G., Petrov, M., Khlaaf, H., Sastry, G., Mishkin, P., Chan, B., Gray, S., . . . Zaremba, W. (2021). *Evaluating Large Language Models Trained on Code (Version 2)*. ArXiv. https://doi.org/10.48550/arxiv.2107.03374

Clark, K. & McCabe, F. (2004). Go! – A Multi-Paradigm Programming Language for Implementing Multi-Threaded Agents. *Annals Of Mathematics And Artificial Intelligence*, *41*(2–4), 171–206. https://doi.org/10.1023/b:amai.0000031195.87297.d9

Coello, C. A. C., Al-Imam, M. & Kouatly, R. (2024). Effectiveness of ChatGPT in Coding: A Comparative Analysis of Popular Large Language Models. *Digital*, *4*(1), 114–125. https://doi.org/10.3390/digital4010005

*Community*. (o. D.). The Scala Programming Language. Abgerufen am 19. März 2024, von https://www.scala-lang.org/community/

Cooper, P. (2009). *Beginning Ruby: From Novice to Professional* (2. Aufl.). Apress. https://doi.org/10.1007/978-1-4302-2364-1

Cravey, D. (2012, August). *C++ - Functional-Style Programming in C++*. Microsoft.com. Abgerufen am 19. März 2024, von https://learn.microsoft.com/en-us/archive/msdn-magazine/2012/august/c-functional-style-programming-in-c

*EvalPlus leaderboard*. (o. D.). Abgerufen am 6. Juni 2024, von https://evalplus.github.io/leaderboard.html

Felleisen, M., Findler, R. B., Flatt, M., Krishnamurthi, S., Barzilay, E., McCarthy, J. & Tobin-Hochstadt, S. (2015). The racket manifesto. *1st Summit On Advances in Programming Languages (SNAPL 2015)*, *32*, 113–128. https://doi.org/10.4230/lipics.snapl.2015.113

Flanagan, D. & Matsumoto, Y. (2008). *The Ruby Programming Language: Everything You Need to Know.* (1. Aufl.). O'Reilly Media, Inc.

Gilbert, L. (2024). *Leetcode-Gym*. GitHub. Abgerufen am 18. Juni 2024, von https://github.com/DieserLaurenz/Leetcode-Gym

Goodwill, J. & Matlock, W. (2015). The Swift programming language. In *Beginning Swift Games Development for iOS* (S. 219–244). Apress. https://doi.org/10.1007/978-1-4842-0400-9_17

Hunt, J. (2023). *Advanced Guide to Python 3 Programming* (2. Aufl.). Springer Cham. https://doi.org/10.1007/978-3-031-40336-1

IBM. (o. D.). *What are Large Language Models (LLMs)? | IBM*. Abgerufen am 12. Juni 2024, von https://www.ibm.com/topics/large-language-models

Jansen, R. H., Vane, V. & De Wolff, I. G. (2016). *TypeScript: Modern JavaScript Development*. Packt Publishing Ltd.

Kelbert, P., Siebert, J. & Jöckel, L. (2024, 12. Dezember). Was sind Large Language Models? Und was ist bei der Nutzung von KI-Sprachmodellen zu beachten? *Fraunhofer IESE*. Abgerufen am 12. Juni 2024, von https://www.iese.fraunhofer.de/blog/large-language-models-ki-sprachmodelle

Kernighan, B. W. & Ritchie, D. M. (1988). *The C programming language* (2. Aufl.). Prentice Hall Professional Technical Reference. https://usuaris.tinet.cat/bertolin/pdfs/c_programming_language.pdf

Klabnik, S. & Nichols, C. (2023). *The Rust Programming Language* (2. Aufl.). No Starch Press.

Kopec, D. (2014). *Dart for Absolute Beginners*. Apress.

Lai, Y., Li, C., Wang, Y., Zhang, T., Zhong, R., Zettlemoyer, L., Yih, S. W., Fried, D., Wang, S. & Yu, T. (2023). DS-1000: A Natural and Reliable Benchmark for Data Science Code Generation. In *International Conference on Machine Learning* (S. 18319–18345). PMLR. https://doi.org/10.48550/arxiv.2211.11501

LeetCode. (o. D.-a). *Account Balance After Rounded Purchase*. leetcode.com. Abgerufen am 14. Juni 2024, von https://leetcode.com/problems/account-balance-after-rounded-purchase/description/

LeetCode. (o. D.-b). *LeetCode QuickStart Guide*. leetcode.com. Abgerufen am 11. Juni 2024, von

https://support.leetcode.com/hc/en-us/articles/360012067053-LeetCode-QuickStart-Guide

LeetCode. (o. D.-c). *Problems - LeetCode*. Abgerufen am 19. März 2024, von

https://leetcode.com/problemset/

LeetCode. (o. D.-d). *Start your Coding Practice*. leetcode.com. Abgerufen am 11. Juni 2024, von

https://support.leetcode.com/hc/en-us/articles/360012016874-Start-your-Coding-Practice

LeetCode. (2024). *What are the environments for the programming languages?* Abgerufen am 21. Januar 2024, von

https://support.leetcode.com/hc/en-us/articles/360011833974-What-are-the-environments-for-the-programming-languages

Liberty, J. (2005). *Programming C#: Building .NET Applications with C#* (4. Aufl.). O'Reilly Media.

Liu, J., Xia, C. S., Wang, Y. & Zhang, L. (2023). *Is Your Code Generated by ChatGPT Really Correct? Rigorous Evaluation of Large Language Models for Code Generation*. arXiv. https://doi.org/10.48550/arxiv.2305.01210

Meta. (o. D.). *Meta Llama 3*. llama.meta.com. Abgerufen am 12. Juni 2024, von

https://llama.meta.com/llama3/

O'Grady, S. (2024, 8. März). *The RedMonk Programming Language rankings: January 2024*. Redmonk. Abgerufen am 6. Juni 2024, von https://redmonk.com/sogrady/2024/03/08/language-rankings-1-24/

OpenAI. (o. D.-a). *How ChatGPT and our language models are developed*. help.openai.com. Abgerufen am 8. Juni 2024, von https://help.openai.com/en/articles/7842364-how-chatgpt-and-our-language-models-are-developed

OpenAI. (o. D.-b). *How your data is used to improve model performance*. help.openai.com. Abgerufen am 2. Juni 2024, von https://help.openai.com/en/articles/5722486-how-your-data-is-used-to-improve-model-performance

OpenAI. (2023a, März 23). *ChatGPT plugins*. openai.com. Abgerufen am 2. Juni 2024, von https://openai.com/index/chatgpt-plugins/

OpenAI. (2023b, November 6). *Introducing GPTs*. openai.com. Abgerufen am 14. Februar 2024, von https://openai.com/blog/introducing-gpts

OpenAI, Achiam, J., Adler, S., Agarwal, S., Ahmad, L., Akkaya, I., Aleman, F. L., Almeida, D., Altenschmidt, J., Altman, S., Anadkat, S., Avila, R., Babuschkin, I., Balaji, S., Balcom, V., Baltescu, P., Bao, H., Bavarian, M., Belgum, J., . . . Zoph, B. (2024). GPT-4 Technical Report (Version 6). *arXiv Preprint*. https://doi.org/10.48550/arXiv.2303.08774

*Other Builtins (Using the GNU Compiler Collection (GCC))*. (o. D.). gcc.gnu.org. Abgerufen am 15. Juni 2024, von https://gcc.gnu.org/onlinedocs/gcc/Other-Builtins.html

Oxley, T., Rajlich, N., Holowaychuk, T. & Young, A. (2017). *Node.Js in action*. Simon and Schuster.

Page, M. J., Moher, D., Bossuyt, P. M., Boutron, I., Hoffmann, T., Mulrow, C. D., Shamseer, L., Tetzlaff, J., Akl, E. A., Brennan, S., Chou, R., Glanville, J., Grimshaw, J., Hróbjartsson, A., Lalu, M. M., Li, T., Loder, E., Mayo‑Wilson, E., McDonald, S., . . . McKenzie, J. E. (2021). PRISMA 2020 explanation and elaboration: updated guidance and exemplars for reporting systematic reviews. *BMJ (Clinical Research Ed.)*, n160. https://doi.org/10.1136/bmj.n160

Pichai, S. & Hassabis, D. (2024, 15. Februar). Our next-generation model: Gemini 1.5. *blog.google*. Abgerufen am 12. Juni 2024, von https://blog.google/technology/ai/google-gemini-next-generation-model-february-2024/

Prettyman, S. (2020). Learn PHP 8: Using MySQL, JavaScript, CSS3, and HTML5. In *Apress eBooks* (2. Aufl.). Apress. https://doi.org/10.1007/978-1-4842-6240-5

Price, M. J. (2019). *C# 8.0 and. NET Core 3.0–Modern Cross-Platform Development: Build applications with C#,. NET Core, Entity Framework Core, ASP. NET Core, and ML. NET using Visual Studio Code*. Packt Publishing Ltd.

Python Software Foundation. (o. D.-a). *General Python FAQ — Python 2.7.18 documentation*. Abgerufen am 19. März 2024, von https://docs.python.org/2.7/faq/general.html

Python Software Foundation. (o. D.-b). *General Python FAQ --- Python 3.11.8 documentation*. Abgerufen am 19. März 2024, von https://docs.python.org/3.11/faq/general.html

Reboucas, M., Pinto, G., Ebert, F., Torres, W., Serebrenik, A. & Castor, F. (2016). An Empirical Study on the Usage of the Swift Programming Language. In *2016 IEEE 23rd international conference on software analysis, evolution, and reengineering (SANER)* (Bd. 1, S. 634–638). IEEE. https://doi.org/10.1109/saner.2016.66

Ritchie, D. M. (1993). The development of the C language. *ACM SIGPLAN Notices*, *28*(3), 201–208. https://doi.org/10.1145/154766.155580

Similarweb. (2024a, Februar). *Traffic-Analysen, Ranking und Publikum von codesignal.com [Februar 2024] | Similarweb*. Abgerufen am 19. März 2024, von https://www.similarweb.com/de/website/codesignal.com/

Similarweb. (2024b, Februar). *Traffic-Analysen, Ranking und Publikum von codewars.com [Februar 2024] | Similarweb*. Abgerufen am 19. März 2024, von https://www.similarweb.com/de/website/codewars.com/

Similarweb. (2024c, Februar). *Traffic-Analysen, Ranking und Publikum von hackerrank.com [Februar 2024] | Similarweb*. Abgerufen am 19. März 2024, von https://www.similarweb.com/de/website/hackerrank.com/

Similarweb. (2024d, Februar). *Traffic-Analysen, Ranking und Publikum von leetcode.com [Februar 2024] | Similarweb*. Abgerufen am 19. März 2024, von https://www.similarweb.com/de/website/leetcode.com/

Similarweb. (2024e, Februar). *Traffic-Analysen, Ranking und Publikum von spoj.com [Februar 2024] | Similarweb*. Abgerufen am 19. März 2024, von https://www.similarweb.com/de/website/spoj.com/

*Stack Overflow Developer Survey 2023*. (2023). Stack Overflow. Abgerufen am 20. März 2024, von https://survey.stackoverflow.co/2023/

Stroustrup, B. (1999). An Overview of the C++ Programming Language. In *The Handbook of Object Technology* (S. 72–94). CRC Press LLC. https://www.stroustrup.com/crc.pdf

*Sunsetting Python 2*. (o. D.). Python. Abgerufen am 19. März 2024, von https://www.python.org/doc/sunset-python-2/

The Elixir Team. (o. D.-a). *Development & Team*. The Elixir Programming Language. Abgerufen am 19. März 2024, von https://elixir-lang.org/development.html

The Elixir Team. (o. D.-b). *The Elixir programming language*. The Elixir Programming Language. Abgerufen am 19. März 2024, von https://elixir-lang.org/

*The Scala Programming Language*. (o. D.). Scala-lang.org. Abgerufen am 19. März 2024, von https://www.scala-lang.org/

Vaswani, A., Shazeer, N., Parmar, N., Uszkoreit, J., Jones, L., Gomez, A. N., Kaiser, L. & Polosukhin, I. (2017). Attention is All you Need. *Advances in Neural Information Processing Systems*, *30*. https://doi.org/10.48550/arXiv.1706.03762

*Why does TypeScript exist?* (o. D.). Typescriptlang.org. Abgerufen am 19. März 2024, von https://www.typescriptlang.org/why-create-typescript

*Why Teach Kotlin*. (o. D.). Kotlinlang.org. Abgerufen am 19. März 2024, von https://kotlinlang.org/education/why-teach-kotlin.html

Williams, A. (2021, 8. Februar). *Rust Foundation*. Hello World! Abgerufen am 19. März 2024, von https://foundation.rust-lang.org/news/2021-02-08-hello-world/

Wilton, P. & McPeak, J. (2010). *Beginning JavaScript* (4. Aufl.). Wiley Publishing.

Yu, H., Shen, B., Ran, D., Zhang, J., Zhang, Q., Ma, Y., Liang, G., Li, Y., Wang, Q. & Xie, T. (2024). CoderEval: A Benchmark of Pragmatic Code Generation with Generative

Pre-trained Models. In *Proceedings of the 46th IEEE/ACM International Conference on Software Engineering* (S. 1–12). https://doi.org/10.1145/3597503.3623316

Zhang, Z., Wen, L., Zhang, S., Chen, D. & Jiang, Y. (2023). Evaluating GPT's Programming Capability through CodeWars' Katas. *arXiv Preprint arXiv:2306.01784*. https://doi.org/10.48550/arxiv.2306.01784

# 8 Anhang

**Tabelle 1**

*Kurzbeschreibungen der 19 Programmiersprachen*

| Name | Beschreibung | Paradigma | Version |
| --- | --- | --- | --- |
| Kotlin | Kotlin ist eine von JetBrains entwickelte Programmiersprache mit statischer Typisierung, die Multiplattform-Fähigkeiten und vollständige Java-Interoperabilität bietet. Sie legt besonderen Wert auf Sicherheit, Modernität und Prägnanz. Kotlin ist die bevorzugte Sprache für die Entwicklung von Android-Apps und wird zudem für Web- und Server-Programmierung sowie Data Science verwendet (*Why Teach Kotlin*, o. D.). | Multiparadigmatisch (primär funktional, imperativ, objektorientiert, prozedural) (*Why Teach Kotlin*, o. D.) | Kotlin 1.9.0 |
| Java | Java ist eine von der Oracle Corporation (ursprünglich Sun Microsystems) entwickelte Programmiersprache mit statischer Typisierung, die Portabilität, Sicherheit und Robustheit in den Vordergrund stellt. Sie ist bekannt für ihre WORA-Philosophie (Write Once, Run Anywhere), was bedeutet, dass einmal geschriebener und kompilierter Code auf jeder Plattform lauffähig ist, die über eine Java Virtual Machine (JVM) verfügt (Arnold et al., 2005). | Objektorientiert (Arnold et al., 2005) | Java 21 |
| Rust | Rust ist eine ursprünglich von Mozilla entwickelte und derzeit von der Rust Foundation betreute Programmiersprache mit statischer Typisierung, die sich auf Sicherheit, Leistung und Produktivität fokussiert (Williams, 2021; Klabnik & Nichols, 2023). In Rust wird der Garbage Collector durch ein System von Ownership, Borrowing und Lifetimes ersetzt. Die Sprache eignet sich für die System- und Anwendungsprogrammierung mit hohen Anforderungen an die Speichersicherheit (Klabnik & Nichols, 2023). | Multiparadigmatisch (Klabnik & Nichols, 2023) | 1.74.1 |
| Scala | Scala ist eine von Martin Odersky und seinem Team an der École Polytechnique Fédérale de Lausanne entwickelte statisch typisierte Programmiersprache mit Fokus auf Skalierbarkeit, Sicherheit und Ausdrucksstärke (*Community*, o. D.; *The Scala Programming Language*, o. D.). Sie vereint das funktionale und objektorientierte Programmierparadigma und eignet sich unter anderem für serverseitige Services, Datenverarbeitung und Frontend-Web-Entwicklung (*The Scala Programming Language*, o. D.). | Funktional, objektorientiert (*The Scala Programming Language*, o. D.) | Scala 2.13.7 |
| Go | Go ist eine von Google entwickelte, statisch typisierte Programmiersprache mit Fokus auf Einfachheit und Performanz. Sie besitzt eine umfangreiche Standardbibliothek und Goroutines für nebenläufige Aufgaben und wird unter anderem für Webserver, verteilte Systeme und Complex Event Processing (CEP) verwendet (Balbaert, 2012). | Multiparadigmatisch (Clark & McCabe, 2004) | Go 1.21 |
| C# | C# ist eine von Microsoft entwickelte statisch typisierte Programmiersprache mit Fokus auf Sicherheit, Einfachheit und Modernität. Sie wurde speziell für das .NET-Framework entwickelt und wird unter anderem für Webservices sowie Desktop- und Mobile-Applikationen verwendet (Price, 2019). | Multiparadigmatisch (primär objektorientiert) (Price, 2019) | C# 12 mit .NET 8 runtime |
| C++ | C++ ist eine von Bjarne Stroustrup bei AT&T entwickelte statisch typisierte Programmiersprache, die C hauptsächlich um Objektorientierung, Datenabstraktion und generische Programmierung erweitert. Sie unterstützt Programmierung auf niedrigen und hohen Abstraktionsebenen und wird unter anderem für Systemprogrammierung verwendet (Stroustrup, 1999). | Multiparadigmatisch (primär prozedural, objektorientiert, generisch) (Cravey, 2012) | C++ 20 |
| PHP | PHP ist eine von Rasmus Lerdorf für die serverseitige Webentwicklung entwickelte dynamisch typisierte Skriptsprache. Sie zielt darauf ab, flexibel, schnell und pragmatisch zu sein, ist in HTML integrierbar und besitzt eine umfassende Kompatibilität zu vielen Datenbanken (Prettyman, 2020). | Multiparadigmatisch (primär objektorientiert, prozedural) (Prettyman, 2020) | PHP 8.2 |

| Name | Beschreibung | Paradigma | Version |
|---|---|---|---|
| Python 2 | Python 2 ist eine von der Python Software Foundation und Guido van Rossum entwickelte, jedoch seit dem 1. Januar 2020 zugunsten von Python 3 eingestellte, dynamisch typisierte Programmiersprache (*Sunsetting Python 2*, o. D.; Hunt, 2023). Sie wird noch immer in einigen Systemen verwendet (Hunt, 2023). | Multiparadigmatisch (primär objektorientiert, prozedural) (Python Software Foundation, o. D.-a) | Python 2.7.12 |
| Python 3 | Python 3 ist die aktuelle Python-Version mit Verbesserungen und Inkompatibilitäten gegenüber Python 2. Sie zielt darauf ab, flexibel und einfach zu sein, ist plattformübergreifend einsetzbar und wird häufig für DevOps-Skripte, Data Science und Machine Learning verwendet (Hunt, 2023). | Multiparadigmatisch (primär objektorientiert, prozedural) (Python Software Foundation, o. D.-b) | Python 3.11 |
| JavaScript | JavaScript ist eine von Brendan Eich bei Netscape entwickelte, von der ECMA standardisierte, dynamisch typisierte Skriptsprache. Sie wird hauptsächlich dazu verwendet, um Webseiten clientseitig interaktiv zu machen (Wilton & McPeak, 2010). Durch die Einführung von Laufzeitumgebungen wie Node.js kann JavaScript auch serverseitig und für andere Zwecke außerhalb der Webentwicklung genutzt werden (Oxley et al., 2017). | Multiparadigmatisch (primär eventbasiert, objektorientiert, prozedural, funktional) (Wilton & McPeak, 2010) | Node.js 20.10.0 |
| TypeScript | TypeScript ist ein von Microsoft entwickeltes Superset von JavaScript mit statischer Typisierung. Sie zielt darauf ab, Fehler im Entwicklungsprozess früher erkennen zu lassen und damit die Entwicklung großer Projekte zu erleichtern (*Why Does TypeScript Exist?*, o. D.). | Multiparadigmatisch (primär objektorientiert, funktional, imperativ) (Jansen et al., 2016) | TypeScript 5.1.6, Node.js 20.10.0 |
| Dart | Dart ist eine von Google entwickelte, statisch typisierte Programmiersprache, die als moderne Alternative zu JavaScript für die Webentwicklung und als Vielzwecksprache konzipiert ist. Sie hat eine ähnliche Syntax zu C und fokussiert sich auf eine organisiertere Struktur, bessere Wartbarkeit und höhere Performance im Vergleich zu JavaScript (Belchin & Juberias, 2015). | Multiparadigmatisch (primär objektorientiert) (Kopec, 2014) | Dart 3.2 |
| Swift | Swift ist eine von Apple entwickelte, statisch typisierte Programmiersprache, die hauptsächlich zur Entwicklung von Anwendungen innerhalb des Apple-Ökosystems verwendet wird. Sie legt den Fokus auf Sicherheit, Leistung und Modernität (Goodwill & Matlock, 2015). Swift wird von Apple als Nachfolger von Objective-C betrachtet (Apple Inc., o. D.). | Multiparadigmatisch (primär objektorientiert, funktional, imperativ) (Reboucas et al., 2016) | Swift 5.9 |
| Ruby | Ruby ist eine von Yukihiro Matsumoto entwickelte, dynamisch typisierte Programmiersprache mit Fokus auf Produktivität und Einfachheit (Flanagan & Matsumoto, 2008). Sie wird unter anderem in der Webentwicklung mit dem bekannten Framework *Ruby on Rails* eingesetzt (Cooper, 2009). | Multiparadigmatisch (primär objektorientiert) (Flanagan & Matsumoto, 2008) | Ruby 3.2 |
| C | C ist eine von Dennis Ritchie bei den Bell Labs entwickelte, statisch typisierte Programmiersprache, die bekannt für ihre Fähigkeit zur hardwarenahen Programmierung ist. Sie wurde ursprünglich für das UNIX-Betriebssystem entwickelt und wird vielfältig angewandt, unter anderem in der System- sowie Anwendungsprogrammierung (Kernighan & Ritchie, 1988). | Imperativ, prozedural (Kernighan & Ritchie, 1988) | gcc 11 |
| Racket | Racket ist eine von PLT Inc. entwickelte Programmiersprache aus der Lisp-Scheme-Familie. Sie unterstützt vorwiegend dynamische, aber auch statische Typisierung und wird vor allem zur Erstellung neuer Programmiersprachen und in der Ausbildung verwendet (Felleisen et al., 2015). | Multiparadigmatisch (primär funktional) (Felleisen et al., 2015) | Racket CS v8.11 |

| Name | Beschreibung | Paradigma | Version |
|---|---|---|---|
| Elixir | Elixir ist eine von José Valim entwickelte dynamisch typisierte Programmiersprache, die auf der Erlang VM basiert (The Elixir Team, o. D.-a; The Elixir Team, o. D.-b). Sie ist konzipiert für wartbare, skalierbare, verteilte und fehlertolerante Systeme mit niedriger Latenz. Elixir wird unter anderem in der Webentwicklung, für Machine Learning und eingebettete Systeme verwendet (The Elixir Team, o. J.-b). | Funktional (The Elixir Team, o. D.-b) | Elixir 1.15 mit Erlang/ OTP 26 |
| Erlang | Erlang ist eine von Ericsson entwickelte dynamisch typisierte Programmiersprache mit Fokus auf Fehlertoleranz, Verfügbarkeit und Nebenläufigkeit. Sie wird in hochverfügbaren Systemen wie Datenbanken, Telekommunikationssystemen und Webdiensten eingesetzt (Armstrong, 2007). | Funktional, nebenläufig (Armstrong, 2007) | Erlang/ OTP 26 |

*Anmerkung.* Quelle: Eigene Darstellung, Quelle für die Spalte der Versionen: LeetCode, 2024

# Eidesstattliche Erklärung

Hiermit versichere ich (Laurenz Gilbert) an Eides statt, dass ich die vorliegende Arbeit „Evaluierung der Code-Generierungsfähigkeiten von ChatGPT 4: Eine vergleichende Analyse in 19 Programmiersprachen" selbstständig und nur mit den angegebenen Quellen und Hilfsmitteln (z. B. Nachschlagewerke oder Internet) angefertigt habe. Alle Stellen der Arbeit, die ich aus diesen Quellen und Hilfsmitteln dem Wortlaut oder dem Sinne nach entnommen habe, sind kenntlich gemacht und im Literaturverzeichnis aufgeführt. Weiterhin versichere ich, dass weder ich noch andere diese Arbeit weder in der vorliegenden noch in einer mehr oder weniger abgewandelten Form als Leistungsnachweise in einer anderen Veranstaltung bereits verwendet haben oder noch verwenden werden. Die Arbeit wurde noch nicht veröffentlicht oder einer anderen Prüfungsbehörde vorgelegt.

Die „Richtlinie zur Sicherung guter wissenschaftlicher Praxis für Studierende an der Universität Potsdam (Plagiatsrichtlinie) - Vom 20. Oktober 2010" ist mir bekannt.

Es handelt sich bei dieser Arbeit um meinen ersten Versuch.

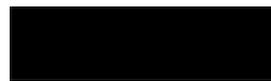

| Berlin, 18.06.2024 | |
|---|---|
| Ort, Datum | Unterschrift |

# Einverständniserklärung

Ich erkläre mich damit einverstanden, meine schriftliche Bachelorarbeit in elektronischer Form einzureichen. Ich bin damit einverstanden, dass die Arbeit mit Hilfe einer Plagiaterkennungssoftware einer Überprüfung unterzogen werden kann.

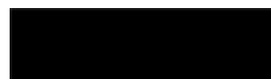

| Berlin, 18.06.2024 | |
|---|---|
| Ort, Datum | Unterschrift |